\documentstyle[12pt]{article}
\title{\bf A Supersymmetric $Sp_{L} \times U_{Y}(1) Model$}
\author{\bf G. G. Blado\\
         Division of Science and Mathematics\\
         University of Minnesota, Morris\\
         Morris, MN 56267\\
         \\
         and \\
         \\
          T. K. Kuo \\
Department of Physics\\
 Purdue University \\
West Lafayette, IN 47907}
\date{ }
\begin{document}
\maketitle
\begin{abstract}
A supersymmetric $Sp_{L}(6) \times U_{Y}(1)$ model (SUSY $Sp(6)$) is proposed
as an extension of the standard electroweak model.
The model is applied in a phenomenological study of $B^{0}_{d} \bar{B}^{0}_{d}$
mixing.
 It is found that the supersymmetric (SUSY) partner
$\tilde{z}^{\prime}$ of the extra $Z^{\prime}$ can significantly cancel the
other
contributions to bring the mixing parameter $x_{d}$ within the experimentally
allowed range $0.57 \stackrel{<}{\sim} x_{d} \stackrel{<}{\sim} 0.77$
for a top mass of $158 \stackrel{<}{\sim} m_{t} \stackrel{<}{\sim} 194$ Gev.
Other
interesting and possibly novel features of flavor changing neutral currents
(FCNC) in
 SUSY theories with horizontal
gauge symmetries are pointed out.
\vspace{.5in}
\begin{tabbing}
PACS 11.30.Pb - Supersymmetry \\
PACS 12.15.Ji - Applications of electroweak models to specific processes \\
PACS 12.15.Mm - Neutral currents
\end{tabbing}
\end{abstract}
\clearpage
\section{Introduction}
 Flavor changing neutral currents (FCNC) pose very stringent tests to
extensions of the standard model (SM). They can severely limit and sometimes
rule out these models (as in the case of some technicolor  models).

One of the most elegant and interesting class of extensions of the standard
model are SUSY models. The recent result \cite{Langacker} (among other things)
 of the unification of the couplings in a SUSY GUT but not in an ordinary GUT
has renewed the confidence of theorists in SUSY models. Needless to say, SUSY
theories have had no serious problems with FCNC's.

Motivated by the viability of SUSY models and the successful phenomenological
studies that have been done on the $Sp_{L}(6) \times U_{Y}(1)$ model
\cite{Blad-Kuo}, \cite{sp6-phenom}, we propose a SUSY $Sp(6)$ model. The
model is developed in analogy to the formulation of the minimal
supersymmetric standard model (MSSM) from the SM \cite{love}, \cite{Rosiek}.

In section \ref{susy-sp1}, we give a brief introduction of the $Sp(6)$ model.
We then proceed to
supersymmetrize this model by writing down the particle spectrum and a workable
supersymmetric $SU_{C}(3) \times Sp_{L}(6) \times U_{Y}(1)$ gauge invariant
lagrangian.

Section \ref{bbmix} examines the phenomenological consequence of SUSY $Sp(6)$
on $B^{0}_{d} \bar{B}^{0}_{d}$ mixing.
Since the gluino contributions in the MSSM still hold in SUSY $Sp(6)$, we
discuss these first. We carefully discuss
the assumptions made and the renormalization-group-modified rotation matrices
of the squark fields. We then give the
explicit expression for the mixing parameter $x_{d}$ in SUSY $Sp(6)$. A plot of
$x_{d}$ versus the top mass $m_{t}$ is
made comparing the SM, MSSM, and the SUSY $Sp(6)$ results. It is found that the
SM and the MSSM may be a bit too high
for $0.57 \stackrel{<}{\sim} x_{d} \stackrel{<}{\sim} 0.77$ with a top mass
$158 \stackrel{<}{\sim} m_{t} \stackrel{<}{\sim} 194$ Gev.
SUSY $Sp(6)$ however, introduces a \underline{cancellation} due to the
$\tilde{z}^{\prime}$ which makes $x_{d}$ fall relatively
well within the experimentally allowable range for a large top mass. A
discussion of further implications follows.

Section \ref{conclude} gives our conclusions and outlook.

\clearpage
\section{The Supersymmetric $SU_{C}(3) \times Sp_{L}(6) \times U_{Y}(1)$ Model}
\label{susy-sp1}

The $SU_{C}(3) \times Sp_{L}(6) \times U_{Y}(1)$ model ($Sp(6)$ model) was
proposed in 1984 \cite{Kuo-Nak} to
address the generation problem of particle physics. A common approach to
introduce a ``horizontal'' group
to tackle the generation problem is the formation of the gauge group
\[G \times SU_{C}(3) \times SU_{L}(2) \times U_{Y}(1)\]
where $G$ is the horizontal group and $SU_{C}(3) \times SU_{L}(2) \times
U_{Y}(1)$ is the familiar SM. This,
however, is not very appealing since it increases the arbitrariness of the
theory by adding another gauge coupling
due to $G$. In addition, these models tend to necessitate the introduction of
more fermions for anomaly cancellation
which may also be questionable since experiment indicates the existence of only
three light fermion families.

The $Sp(6)$ group, however, has the unique feature of unifying $G$ and
$SU_{L}(2)$ into a single horizontal group
$Sp_{L}(6)$ without introducing extra fermions into the theory for anomaly
cancellation since $Sp(6)$ is anomaly-free.
$Sp(6)$ has a horizontal subgroup $SU_{H}(3)$ which mixes the different
generations. In this extension, $Sp(6)$ has
a six dimensional representation \underline{6} , which contains the six leptons
and quarks in a multiplet. $Sp(6)$
decomposes into three \underline{2} of $SU(2)$ which gives rise to the doublets
in the three generations of leptons
and quarks.

The Lie group  $Sp(6)$ has 21 generators, $T^{(i)},\;i = 1,\,2,\, \ldots,\, 21$
whose $6 \times 6$ representation
are given by \cite{Blad-Kuo},
\begin{eqnarray}
\frac{1}{2}\sqrt{\frac{1}{2}}\;(\sigma_{1},\sigma_{2},\sigma_{3})
\otimes \lambda_{S}^{i} & , &
\lambda_{S}^{i}=\lambda^{0},\lambda^{1},\lambda^{3},\lambda^{4}
,\lambda^{6},\lambda^{8}
\label{eq1a*p*2-2}
 \\
\frac{1}{2}\sqrt{\frac{1}{2}}\;{\bf 1}\otimes \lambda_{A}^{i}
                             & , &
\lambda_{A}^{i}=\lambda^{2},\lambda^{5},\lambda^{7}
\label{eq1b*p*2-2}
\end{eqnarray}
where the $\sigma_{i},\;\;i=1,2,3$ and $\lambda^{i},\;\;i=0,1,\ldots,8$
are the Pauli and Gell-Mann matrices respectively. For future reference we
assign arbitrarily $T^{(i)}$ to equations
\ref{eq1a*p*2-2} and \ref{eq1b*p*2-2} in equation \ref{eq3*p*ApA-1} of the
appendix . The normalization
in equations \ref{eq1a*p*2-2} and \ref{eq1b*p*2-2} are such that
\begin{equation}
Tr(T^{(i)}T^{(j)})=\frac{1}{2}\delta^{ij}\;\;\;\;i\:,\:j\:=1,2,\ldots,21\;.
\end{equation}
The three $SU(2)$ subgroups (which we denote by $SU_{i^{\prime}}(2),\;
i^{\prime} = 1,\,2,\,3$) to which $Sp(6)$
decomposes have the following generators
\begin{equation}
\vec{\Sigma}_{1} \equiv \frac{1}{2}\vec{\sigma}\otimes\left( \begin{array}{ccc}
1 &   &   \\
  & 0 &   \\
  &   & 0
\end{array} \right)
\;\;;\;\;
\vec{\Sigma}_{2} \equiv \frac{1}{2}\vec{\sigma}\otimes\left( \begin{array}{ccc}
0 &   &   \\
  & 1 &   \\
  &   & 0
\end{array} \right)
\;\;;\;\;
\vec{\Sigma}_{3} \equiv \frac{1}{2}\vec{\sigma}\otimes\left( \begin{array}{ccc}
0 &   &   \\
  & 0 &   \\
  &   & 1
\end{array} \right)\;.
\label{eq3*p*2-2}
\end{equation}
In the symmetry breaking scheme,
\begin{eqnarray}
Sp(6) & \longrightarrow & SU_{1}(2)\times SU_{2}(2)\times SU_{3}(2)\nonumber \\
      & \longrightarrow & SU_{12}(2)\times SU_{3}(2)\nonumber \\
      & \longrightarrow & SU_{123}(2) = SU_{L}(2)
\label{eq4*p*2-2}
\end{eqnarray}
$SU_{12}(2)$ and $SU_{123}(2)$ are diagonal $SU(2)$ subgroups of the relevant
direct product groups. As indicated in relation
\ref{eq4*p*2-2}, the group $SU_{123}(2)$ is to be identified with the
$SU_{L}(2)$ of the SM. If we denote (with space-time indices
suppressed) $\vec{A} \equiv (A_{1}, A_{2}, A_{3})$ to be the $SU_{L}(2)$ gauge
bosons, and $\vec{A}^{(i^{\prime})}$ with
$i^{\prime} = 1,\,2,\,3$ to be the gauge bosons associated with the three
$SU_{i^{\prime}}(2)$ subgroups in the symmetry breaking
scheme of relation \ref{eq4*p*2-2}, we have
\begin{equation}
\vec{A} = \frac{1}{\sqrt{3}} \left( \vec{A}^{(1)} + \vec{A}^{(2)} +
\vec{A}^{(3)}  \right)
\label{eq5*p*2-2}
\end{equation}
Equation \ref{eq5*p*2-2} indicates why the $SU_{L}(2)$ gauge bosons couple
universally to the three generations and it implies that
\begin{equation}
g_{2} = \frac{1}{\sqrt{3}} g_{sp}
\end{equation}
where $g_{2}$ and $g_{sp}$ are the $SU_{L}(2)$ and $Sp_{L}(6)$ gauge coupling
constants respectively. The other set of relatively
light new gauge bosons are
\begin{equation}
\left( W_{1}^{\prime},\,W_{2}^{\prime}, Z^{\prime}\right) = \frac{1}{\sqrt{6}}
\left( \vec{A}^{(1)} + \vec{A}^{(2)} - 2 \vec{A}^{(3)}  \right)
\label{eq2a*p*2-3}
\end{equation}
\begin{equation}
\left( W_{1}^{\prime \prime},\,W_{2}^{\prime \prime}, Z^{\prime \prime}\right)
= \frac{1}{\sqrt{2}}
\left( \vec{A}^{(1)} - \vec{A}^{(2)} \right)
\label{eq2b*p*2-3}
\end{equation}
{}From equations \ref{eq2a*p*2-3} and \ref{eq2b*p*2-3}, it is evident that
these gauge bosons \underline{do not}
couple universally to the three generations. The lightest of these extra gauge
bosons which can possibly be detected
in the near future is the neutral gauge boson $Z^{\prime}$.

To get the coupling of the $Z^{\prime}$ with the fermions, we first write the
term in the $Sp(6)$ model lagrangian
describing the kinetic energy of the matter (fermion) fields and gauge-matter
interactions (see the appendix
 for notation, conventions and some relevant formulas).
\begin{eqnarray}
\cal L \mit _{kin}  & = & i \bar{\Psi}^{\prime}_{(i_{6}) L} \gamma_{\mu} \cal D
\mit ^{(2) \mu} \Psi^{\prime}_{(i_{6}) L} +
                         i \rm \bar{\Psi} \mit ^{I}_{rt} \gamma_{\mu} \cal D
\mit ^{(1) \mu} \Psi^{I}_{rt} + \nonumber\\
                   &   & i (\bar{\Psi}^{\prime}_{Q})_{(i_{6}) \alpha L}
\gamma_{\mu} \nabla^{\mu}
                         (\Psi^{\prime}_{Q})_{(i_{6})
                         \alpha L} + i (\bar{\Psi}_{u})^{I}_{\alpha rt}
\gamma_{\mu} \cal D \mit ^{(2) \mu} (\Psi_{u})^{I}_{\alpha rt}
                         + \label{eq3*p*2-3}\\
                   &   & i (\bar{\Psi}_{d})^{I}_{\alpha rt} \gamma_{\mu} \cal D
\mit ^{(2) \mu} (\Psi_{d})^{I}_{\alpha rt} \nonumber
\end{eqnarray}
In this paper, primed fermion (and later sfermion) fields are the initial
fields (as opposed to the physical, mass
eigenstate fields).

As an example of extracting the $Z^{\prime}$-quark interaction terms, let us
rewrite the third term in equation \ref{eq3*p*2-3}.
Using equation
\ref{eq12*p*ApA-1},
\begin{equation}
\begin{array}{ll}
i (\Psi^{\prime}_{Q})_{(i_{6}) \alpha L} \gamma^{\mu} & \left[
\partial^{\mu} (\Psi^{\prime}_{Q})_{(i_{6}) \alpha L}
+ i g_{3} G^{\mu}_{a} Y^{a}_{ \alpha \beta} (\Psi^{\prime}_{Q})_{(i_{6}) \beta
L}
+ i  g_{sp} A^{\mu}_{j} T^{(j)}_{i_{6} j_{6}} (\Psi^{\prime}_{Q})_{(j_{6})
\alpha L}\right .\\
             & +\, \left . i \frac{g_{1}}{2} B^{\mu} y
(\Psi^{\prime}_{Q})_{(i_{6}) \alpha L}
\right]
\end{array}
\end{equation}
Looking at the term
\begin{equation}
i (\bar{\Psi}^{\prime}_{Q})_{(i_{6}) \alpha L} \gamma_{\mu}
\left[ i g_{sp} A^{\mu}_{j} T^{(j)}_{i_{6} j_{6}} (\Psi^{\prime}_{Q})_{(j_{6})
\alpha L} \right]
\label{eq1a*p*2-4}
\end{equation}
we have to identify which $A^{\mu}_{j}$ will correspond to $Z^{\prime}$ (or
possibly a linear combination of $A^{\mu}_{j}$).
{}From equation \ref{eq2a*p*2-3},
\begin{equation}
Z^{\prime} = \frac{1}{\sqrt{6}}
 \left(
\vec{A}^{(1)}_{3} + \vec{A}^{(2)}_{3} - 2 \vec{A}^{(3)}_{3}
\right)
\label{eq2*p*2-4}
\end{equation}
We then have to get the relation of $\left( \vec{A}^{(1)} \right)^{\mu}, \left(
\vec{A}^{(2)} \right)^{\mu}, \left( \vec{A}^{(3)} \right)^{\mu}$ with
$A^{\mu}_{j}$ of $Sp(6)$. The key point to realize here is that these three
sets of gauge bosons associated with the
three $SU_{(i^{\prime})}(2)$ groups would then correspond to the three sets of
generators in equation~\ref{eq3*p*2-2} in the following
manner,
\begin{equation}
\left( \vec{A}^{(i^{\prime})} \right)^{\mu} \longleftrightarrow
\vec{\Sigma}_{i^{\prime}},\; i^{\prime} = 1,2,3
\label{eq3*p*2-4}
\end{equation}
and also
\begin{equation}
A^{\mu}_{j} \longleftrightarrow T^{(j)}.
\label{eq4*p*2-4}
\end{equation}
Hence from equations \ref{eq3*p*2-4} and \ref{eq4*p*2-4},  if we can write
$T^{(j)}$ as linear combinations of
$\vec{\Sigma}_{i^{\prime}}$, then we could write $A^{\mu}_{j}$ as linear
combinations of $\left( \vec{A}^{(i^{\prime})} \right)^{\mu}$
and vice versa. Knowing the expressions of $\left( \vec{A}^{(i^{\prime})}
\right)^{\mu}$ in terms of the $A^{\mu}_{j}$ of $Sp(6)$, we can
then put these expressions into \ref{eq2*p*2-4}. Going through these steps, we
find
\begin{equation}
\label{eq5*p*2-4}
Z^{\prime}_{\nu} = A_{\nu (18)}
\end{equation}
where (see also the appendix  equation \ref{eq3*p*ApA-1})
\begin{equation}
\mit A_{\nu (18)} \longleftrightarrow T^{(18)} = \frac{1}{2 \sqrt{2}}
\sigma_{3} \otimes \lambda^{8}
\end{equation}
Hence, to get the term in equation \ref{eq1a*p*2-4} describing the interaction
of the $Z^{\prime}$ with the left handed quarks
we look at the term
\begin{equation}
i (\bar{\Psi}^{\prime}_{Q})_{(i_{6}) \alpha L} \gamma_{\mu}
\left[ i g_{sp} A^{\mu}_{(18)} T^{(18)}_{i_{6} j_{6}}
(\Psi^{\prime}_{Q})_{(j_{6}) \alpha L} \right] =
i (\bar{\Psi}^{\prime}_{Q})_{(i_{6}) \alpha L} \gamma_{\mu}
\left[ i g_{sp} Z^{\prime \mu} T^{(18)}_{i_{6} j_{6}}
(\Psi^{\prime}_{Q})_{(j_{6}) \alpha L} \right].
\end{equation}
Of course, one then has to express the initial (primed) fields in terms of the
physical fields by rotating the initial fields
using the appropriate unitary matrices.

Let us now describe the \underline{(minimal) SUSY $Sp(6)$ model}. To establish
notation, let us list down the particle spectrum of SUSY
$Sp(6)$ in tables \ref{tbl1*p*2-7}, \ref{tbl1*p*2-8}, \ref{tbl1*p*2-9},
\ref{tbl1*p*2-10} and \ref{tbl2*p*2-10}. We also list down
their quantum numbers in table \ref{tbl1*p*2-11}.
\begin{table}[htbp]
\centering
  \caption{Gauge Bosons and Gauginos in SUSY $Sp(6)$}
\label{tbl1*p*2-7}
  \vspace{0.5cm}
\begin{tabular}{l|c|c|c} \hline
vector & bosonic & fermionic & auxiliary \\
   superfields & components & components & fields \\ \hline
$\cal G \mit ^{a}$ (for $SU_{C}(3)$) & $G^{a}_{\mu}$ & $\lambda^{a}_{G}$ &
$D^{a}_{G}$ \\ \hline
$\cal A \mit ^{i}$ (for $Sp_{L}(6)$) & $A^{i}_{\mu}$ & $\lambda^{i}_{A}$ &
$D^{i}_{A}$ \\ \hline
$\hat{\cal B}$ (for $U_{Y}(1)$) & $B_{\mu}$ & $\lambda_{B}$ & $D_{B}$ \\ \hline
\end{tabular}
\end{table}
\begin{table}[htbp]
\centering
  \caption{Leptons and Sleptons in SUSY $Sp(6)$}
\label{tbl1*p*2-8}
  \vspace{0.5cm}
\begin{tabular}{l|c|c} \hline
chiral & bosonic & fermionic \\
   superfields & components & components \\ \hline
%
$\pounds_{(i_{6})} = \left[ \begin{array}{c}
                      \pounds_{1}  \\
                      \pounds_{2}  \\
                      \pounds_{3}  \\
                      \pounds_{4}  \\
                      \pounds_{5}  \\
                      \pounds_{6}
                     \end{array} \right]$
& 
$
 \left[ \begin{array}{c}
                       L^{\prime}_{1}   \\
                       L^{\prime}_{2}   \\
                       L^{\prime}_{3}   \\
                       L^{\prime}_{4}   \\
                       L^{\prime}_{5}   \\
                       L^{\prime}_{6}
                     \end{array} \right]  =  \left[ \begin{array}{c}
                                           \tilde{\nu}^{\prime}_{e L}\\
                                           \tilde{\nu}^{\prime}_{\mu L}\\
                                           \tilde{\nu}^{\prime}_{\tau L}\\
                                            \tilde{e}^{\prime}_{L}  \\
                                            \tilde{\mu}^{\prime}_{L}  \\
                                            \tilde{\tau}^{\prime}_{L}
                                           \end{array} \right]
$
& 
$
                       \left[ \begin{array}{c}
                       \psi^{\prime}_{l 1}  \\
                       \psi^{\prime}_{l 2}   \\
                       \psi^{\prime}_{l 3}   \\
                       \psi^{\prime}_{l 4}   \\
                       \psi^{\prime}_{l 5}   \\
                       \psi^{\prime}_{l 6}
                     \end{array} \right]  =  \left[ \begin{array}{c}
                                           {\nu}^{\prime}_{e L}\\
                                           {\nu}^{\prime}_{\mu L}\\
                                           {\nu}^{\prime}_{\tau L}\\
                                           {e}^{\prime}_{L} \\
                                           {\mu}^{\prime}_{L} \\
                                           {\tau}^{\prime}_{L}
                                           \end{array} \right]
$
\\ \hline
%
$\cal R \mit ^{1}$
& 
$R^{\prime 1} = \tilde{e}^{\prime + 1}_{R} = \tilde{e}^{\prime +}_{R}$
& 
$\psi^{\prime 1}_{R} = (e^{\prime 1}_{L})^{c} = e^{\prime c}_{L}$
\\ \hline
%
$\cal R \mit ^{2}$
& 
$R^{\prime 2} = \tilde{e}^{\prime + 2}_{R} = \tilde{\mu}^{\prime +}_{R}$
& 
$\psi^{\prime 2}_{R} = (e^{\prime 2}_{L})^{c} = \mu^{\prime c}_{L}$
\\ \hline
%
$\cal R \mit ^{3}$
& 
$R^{\prime 3} = \tilde{e}^{\prime + 3}_{R} = \tilde{\tau}^{\prime +}_{R}$
& 
$\psi^{\prime 3}_{R} = (e^{\prime 3}_{L})^{c} = \tau^{\prime c}_{L}$
\\ \hline
\end{tabular}
\end{table}
\begin{table}[htbp]
\centering
  \caption{Quarks and Squarks in SUSY $Sp(6)$ }
\label{tbl1*p*2-9}
  \vspace{0.5cm}
\begin{tabular}{l|c|c} \hline
chiral & bosonic & fermionic \\
   superfields & components & components \\ \hline
%
$\cal Q \mit _{(i_{2}) \alpha} = \left[ \begin{array}{c}
                      \cal Q \mit _{1 \alpha}  \\
                       \cal Q \mit _{2 \alpha}  \\
                       \cal Q \mit _{3 \alpha}  \\
                       \cal Q \mit _{4 \alpha}  \\
                       \cal Q \mit _{5 \alpha}  \\
                       \cal Q \mit _{6 \alpha}
                     \end{array} \right]$
& 
$
\!\!\!\!
                   \left[ \begin{array}{c}
                       Q^{\prime}_{1 \alpha}  \\
                       Q^{\prime}_{2 \alpha}  \\
                       Q^{\prime}_{3 \alpha}  \\
                       Q^{\prime}_{4 \alpha}  \\
                       Q^{\prime}_{5 \alpha}  \\
                       Q^{\prime}_{6 \alpha}
                     \end{array} \right] = \left[ \begin{array}{c}
                                           \tilde{u}^{\prime}_{L \alpha}\\
                                           \tilde{c}^{\prime}_{L \alpha} \\
                                           \tilde{t}^{\prime}_{L \alpha} \\
                                           \tilde{d}^{\prime}_{L \alpha} \\
                                           \tilde{s}^{\prime}_{L \alpha} \\
                                           \tilde{b}^{\prime}_{L \alpha}
                                           \end{array} \right]
$
& 
$
\!\!\!\!
                  \left[ \begin{array}{c}
                       \psi^{\prime}_{q 1 \alpha}  \\
                       \psi^{\prime}_{q 2 \alpha}   \\
                       \psi^{\prime}_{q 3 \alpha}   \\
                       \psi^{\prime}_{q 4 \alpha}   \\
                       \psi^{\prime}_{q 5 \alpha}   \\
                       \psi^{\prime}_{q 6 \alpha}
                     \end{array} \right] = \left[ \begin{array}{c}
                                           {u}^{\prime}_{L \alpha}\\
                                           {c}^{\prime}_{L \alpha} \\
                                           {t}^{\prime}_{L \alpha} \\
                                           {d}^{\prime}_{L \alpha} \\
                                           {s}^{\prime}_{L \alpha} \\
                                           {b}^{\prime}_{L \alpha}
                                           \end{array} \right]
$
\\ \hline
%
$\cal D \mit ^{1}_{\alpha} = \cal D \mit _{\alpha}$
& 
$D^{\prime 1}_{\alpha} = \tilde{d}^{\prime \ast 1}_{R \alpha} =
\tilde{d}^{\prime \ast}_{R \alpha}$
& 
$\psi^{\prime 1}_{D \alpha} = (d^{\prime 1}_{L \alpha})^{c} = d^{\prime c}_{L
\alpha}$
\\ \hline
%
$\cal D \mit ^{2}_{\alpha} = \cal S \mit _{\alpha}$
& 
$D^{\prime 2}_{\alpha} = \tilde{d}^{\prime \ast 2}_{R \alpha} =
\tilde{s}^{\prime \ast}_{R \alpha}$
& 
$\psi^{\prime 2}_{D \alpha} = (d^{\prime 2}_{L \alpha})^{c} = s^{\prime c}_{L
\alpha}$
\\ \hline
%
$\cal D \mit ^{3}_{\alpha} = \cal B \mit _{\alpha}$
& 
$D^{\prime 3}_{\alpha} = \tilde{d}^{\prime \ast 3}_{R \alpha} =
\tilde{b}^{\prime \ast}_{R \alpha}$
& 
$\psi^{\prime 3}_{D \alpha} = (d^{\prime 3}_{L \alpha})^{c} = b^{\prime c}_{L
\alpha}$
\\ \hline
%
$\cal U \mit ^{1}_{\alpha} = \cal U \mit _{\alpha}$
& 
$U^{\prime 1}_{\alpha} = \tilde{u}^{\prime \ast 1}_{R \alpha} =
\tilde{u}^{\prime \ast}_{R \alpha}$
& 
$\psi^{\prime 1}_{U \alpha} = (u^{\prime 1}_{L \alpha})^{c} = u^{\prime c}_{L
\alpha}$
\\ \hline
%
$\cal U \mit ^{2}_{\alpha} = \cal C \mit _{\alpha}$
& 
$U^{\prime 2}_{\alpha} = \tilde{u}^{\prime \ast 2}_{R \alpha} =
\tilde{c}^{\prime \ast}_{R \alpha}$
& 
$\psi^{\prime 2}_{U \alpha} = (u^{\prime 2}_{L \alpha})^{c} = c^{\prime c}_{L
\alpha}$
\\ \hline
%
$\cal U \mit ^{3}_{\alpha} = \cal T \mit _{\alpha}$
& 
$U^{\prime 3}_{\alpha} = \tilde{u}^{\prime \ast 3}_{R \alpha} =
\tilde{t}^{\prime \ast}_{R \alpha}$
& 
$\psi^{\prime 3}_{U \alpha} = (u^{\prime 3}_{L \alpha})^{c} = t^{\prime c}_{L
\alpha}$
\\ \hline
\end{tabular}
\end{table}
\begin{table}[htbp]
\centering
  \caption{Higgs and Higgsinos in SUSY $Sp(6)$}
\label{tbl1*p*2-10}
  \vspace{0.5cm}
\begin{tabular}{l|c|c} \hline
chiral & bosonic & fermionic \\
    superfields & components & components\\ \hline
%
$ \hat{\Phi}_{(i_{6})} = \left[ \begin{array}{c}
                      \hat{\Phi}_{1}  \\
                      \hat{\Phi}_{2}  \\
                      \hat{\Phi}_{3}  \\
                      \hat{\Phi}_{4}  \\
                      \hat{\Phi}_{5}  \\
                      \hat{\Phi}_{6}
                     \end{array} \right]$
&
$\phi_{(i_{6})}  =  \left[ \begin{array}{c}
                       \phi_{1}  \\
                       \phi_{2}   \\
                       \phi_{3}    \\
                       \phi_{4}   \\
                       \phi_{5}  \\
                       \phi_{6}
                     \end{array} \right]$
& 
$\psi_{\phi\,(i_{6})}  =  \left[ \begin{array}{c}
                       \psi_{\phi 1}  \\
                       \psi_{\phi 2}   \\
                       \psi_{\phi 3} \\
                       \psi_{\phi 4}  \\
                       \psi_{\phi 5}    \\
                       \psi_{\phi 6}
                     \end{array} \right]$ \\ \hline
$\cal H \mit _{(i_{6})} = \left[ \begin{array}{c}
                      \cal H \mit _{1}  \\
                       \cal H \mit _{2}  \\
                       \cal H \mit _{3}  \\
                       \cal H \mit _{4}   \\
                       \cal H \mit _{5}   \\
                       \cal H \mit _{6}
                     \end{array} \right]$
&
$H_{(i_{6})}  =  \left[ \begin{array}{c}
                       H_{1}  \\
                       H_{2}  \\
                       H_{3}  \\
                       H_{4}   \\
                       H_{5}  \\
                       H_{6}
                     \end{array} \right]$
& 
$\psi_{H\,(i_{6})}  =  \left[ \begin{array}{c}
                       \psi_{H 1}  \\
                       \psi_{H 2}   \\
                       \psi_{H 3}   \\
                       \psi_{H 4}   \\
                       \psi_{H 5}   \\
                       \psi_{H 6}
                     \end{array} \right]$ \\ \hline
\end{tabular}
\end{table}
\begin{table}[htbp]
\centering
  \caption{Auxiliary Fields of the Matter Multiplets in SUSY $Sp(6)$}
\label{tbl2*p*2-10}
  \vspace{0.5cm}
\begin{tabular}{l|l} \hline
chiral & auxiliary \\
  superfields & fields \\ \hline
$\pounds_{(i_{6})}$ & $F_{l\,(i_{6})}$ \\ \hline
$\cal R \mit ^{I}$ & $F^{I}_{R}$ \\ \hline
$\cal Q \mit _{(i_{6}) \alpha}$ & $F_{q\, (i_{6}) \alpha}$ \\ \hline
$\cal D \mit ^{I}_{\alpha}$ & $F^{I}_{D \alpha}$ \\ \hline
$\cal U \mit ^{I}_{\alpha}$ & $F^{I}_{U \alpha}$ \\ \hline
$\hat{\Phi}_{(i_{6})}$ & $F_{\phi\, (i_{6})}$ \\ \hline
$\cal H \mit _{(i_{6})}$ & $F_{H\, (i_{6})}$ \\ \hline
\end{tabular}
\end{table}
\begin{table}[htbp]
\centering
  \caption{Quantum Numbers of Particles in SUSY $Sp(6)$}
\label{tbl1*p*2-11}
  \vspace{0.5cm}
\begin{tabular}{l|c|c|c} \hline
 superfields & $SU_{C}(3)$ & $Sp_{L}(6)$ & y \\
  & transformation & transformation & (hypercharge) \\ \hline
$\cal G \mit ^{a}$ & $\underline{8}$ & $\underline{1}$ & $0$ \\ \hline
$\cal A \mit ^{i}$ & $\underline{1}$ & $\underline{21}$ & $0$ \\ \hline
$\hat{\cal B}$ & $\underline{1}$ & $\underline{1}$ & $0$ \\ \hline
$\pounds_{(i_{6})}$ & $\underline{1}$ & $\underline{6}$ & $-1$ \\ \hline
$\cal R \mit ^{I}$ & $\underline{1}$ & $\underline{1}$ & $2$ \\ \hline
$\cal Q \mit _{(i_{6}) \alpha} $ & $\underline{3}$ & $\underline{6}$ &
$\frac{1}{3}$ \\ \hline
$\cal D \mit ^{I}_{\alpha}$ & $\underline{\bar{3}}$ & $\underline{1}$ &
$\frac{2}{3}$ \\ \hline
$\cal U \mit ^{I}_{\alpha}$ & $\underline{\bar{3}}$ & $\underline{1}$ &
$-\frac{4}{3}$ \\ \hline
$\hat{\Phi} \mit _{(i_{6})}$ & $\underline{1}$ & $\underline{6}$ & $1$ \\
\hline
$\cal H \mit _{(i_{6})}$ & $\underline{1}$ & $\underline{6}$ & $-1$ \\ \hline
\end{tabular}
\end{table}

Note that in the tables, the fermionic components are two-component spinors.
For the
chiral superfields, we denote these two-component spinors by the lower case
greek letter $\psi$
(as opposed to the usual four-component Dirac spinor as in equation
\ref{eq3*p*2-3} which we denote by the upper case greek
letter $\Psi$). The complete set of formulas for converting two-component to
four-component spinors are given in equations (A19)
to (A23) of reference \cite{Haber-Kane}. There, $\Psi_{i}$ is defined as
\begin{equation}
\Psi_{i} =  \left( \begin{array}{c}
                       \xi_{i}  \\
                       \bar{\eta}_{i}
                     \end{array} \right)
\end{equation}
where $\xi_{i}$ and $\eta_{i}$ are two-component spinors. These conversions are
indispensable in deriving Feynman rules.

The bosonic superpartners of the ordinary fermions, namely leptons and quarks
are indicated by the same letter but with a tilde
on top (example: if $e \longrightarrow$ electron then $\tilde{e}
\longrightarrow$ selectron).  As usual we refer to superpartners
of fermions as sfermions, while for gauge bosons, we refer to their
superpartners as gauginos.

$\cal G \mit ^{a}$, $\cal A \mit ^{i}$ and $\hat{\cal B}$ are the vector
superfield multiplets of the gluons, $Sp_{L}(6)$, and
$U_{Y}(1)$ gauge bosons respectively and their superpartners.  $\pounds$ and
$\cal Q$ denote the superfield multiplet of left-handed leptons and quarks
respectively and their superpartners while  the $\cal R$,
$\cal D$ and $\cal U$ on the other hand denote the superfield multiplet of
right-handed electron-type leptons, down-type quarks and up-type
quarks respectively and their superpartners. The two types of higgs chiral
superfield multiplets are denoted by $ \hat{\Phi}_{(i_{6})}$
and $\cal H \mit _{(i_{6})}$. As in the MSSM, we
introduce the two types of higgs  to cancel the anomaly due to the superpartner
of the original higgs.

Given the above particle spectrum, we can now write a (minimal) supersymmetric
$SU_{C}(3) \times Sp_{L}(6) \times U_{Y}(1)$
gauge-invariant lagrangian given by
\begin{equation}
\cal L =  \cal L \mit _{YM} + \cal L \mit _{kin} + \cal L \mit
_{superpotential} + \cal L \mit _{soft-breaking}
\end{equation}
where
\begin{equation}
\begin{array}{lll}
\cal L \mit _{YM}\!\! & = \!\!& \frac{1}{4 k_{3} (2 g_{3})^{2}} \rm Tr \mit
\left[  \left.W^{\eta}_{G} W_{G \eta} \right|_{\theta \theta} +
                                                     \left. \rm \bar{\mit W}_{G
\dot{\eta}} \bar{\mit W}^{\dot{\eta}}_{G} \right|_{\rm \bar{\theta}
\bar{\theta}} \right] +\\
                  &   & \frac{1}{4 k (2 g_{sp})^{2}} \rm Tr \mit \left[
\left.W^{\eta}_{A} W_{A \eta} \right|_{\theta \theta} +
                                                     \left. \rm \bar{\mit W}_{A
\dot{\eta}} \bar{\mit W}^{\dot{\eta}}_{A} \right|_{\rm \bar{\theta}
\bar{\theta}} \right] +\\
                  &   & \frac{1}{4} \left[  \left.W^{\eta}_{B} W_{B \eta}
\right|_{\theta \theta} +
                                                     \left. \rm \bar{\mit W}_{B
\dot{\eta}} \bar{\mit W}^{\dot{\eta}}_{B} \right|_{\rm \bar{\theta}
\bar{\theta}} \right]
\end{array}
\label{eq2a*p*2-12}
\end{equation}
\begin{equation}
\begin{array}{lll}
\cal L \mit _{kin} & = & \left. \pounds^{\dagger}_{(i_{6})}
                         e^{2 \left[
                                    g_{sp} T^{(i)}_{i_{6} j_{6}} \cal A \mit
_{i} +
                                    g_{1} \left( \frac{1}{2}\right) \left( -1
\right) \delta_{i_{6} j_{6}} \rm \hat{\cal B}
                             \right]}
                         \pounds_{(j_{6})} \right|_{\theta \theta \rm
\bar{\theta} \bar{\theta}} + \\
                   &   & \left. \cal R \mit ^{I \dagger}
                         e^{2 \left[
                                    g_{1} \left( \frac{1}{2}\right) \left( 2
\right)  \rm \hat{\cal B}
                             \right]}
                         \cal R \mit ^{I} \right|_{\theta \theta \rm
\bar{\theta} \bar{\theta}} + \\
                   &   & \left. \cal Q \mit ^{\dagger}_{(i_{6}) \alpha}
                         e^{2 \left[g_{3} Y^{a}_{\alpha \beta} \cal G \mit _{a}
\delta_{i_{6} j_{6}} +
                                    g_{sp} \delta_{\alpha \beta}
T^{(i)}_{i_{6} j_{6}} \cal A \mit _{i} +
                                    g_{1} \left( \frac{1}{2}\right) \left(
\frac{1}{3} \right) \delta_{\alpha \beta} \delta_{i_{6} j_{6}} \rm \hat{\cal B}
                             \right]}
                         \cal Q \mit _{(j_{6}) \beta} \right|_{\theta \theta
\rm \bar{\theta} \bar{\theta}} + \\
                   &   & \left. \cal D \mit ^{I \dagger}_{\alpha}
                         e^{2 \left[g_{3} \rm \bar{\mit Y}^{a}_{\alpha \beta}
\cal G \mit _{a} +
                                    g_{1} \left( \frac{1}{2}\right) \left(
\frac{2}{3} \right)  \delta_{\alpha \beta} \rm \hat{\cal B}
                             \right]}
                         \cal D \mit ^{I}_{\beta} \right|_{\theta \theta \rm
\bar{\theta} \bar{\theta}} + \\
                   &   & \left. \cal U \mit ^{I \dagger}_{\alpha}
                         e^{2 \left[g_{3} \rm \bar{\mit Y}^{a}_{\alpha \beta}
\cal G \mit _{a} +
                                    g_{1} \left( \frac{1}{2}\right) \left(
-\frac{4}{3} \right)  \delta_{\alpha \beta} \rm \hat{\cal B}
                             \right]}
                         \cal U \mit ^{I}_{\beta} \right|_{\theta \theta \rm
\bar{\theta} \bar{\theta}} + \\
                  &    & \left. \cal H \mit ^{\dagger}_{(i_{6})}
                         e^{2 \left[
                                    g_{sp} T^{(i)}_{i_{6} j_{6}} \cal A \mit
_{i} +
                                    g_{1} \left( \frac{1}{2}\right) \left( -1
\right) \delta_{i_{6} j_{6}} \rm \hat{\cal B}
                             \right]}
                         \cal H \mit _{(j_{6})} \right|_{\theta \theta \rm
\bar{\theta} \bar{\theta}} + \\
                  &    & \left. \hat{\Phi} ^{\dagger}_{(i_{6})}
                         e^{2 \left[
                                    g_{sp} T^{(i)}_{i_{6} j_{6}} \cal A \mit
_{i} +
                                    g_{1} \left( \frac{1}{2}\right) \left( 1
\right) \delta_{i_{6} j_{6}} \rm \hat{\cal B}
                             \right]}
                         \hat{\Phi} _{(j_{6})} \right|_{\theta \theta \rm
\bar{\theta} \bar{\theta}}
\end{array}
\end{equation}
\begin{equation}
\begin{array}{lll}
\cal L \mit _{superpotential} & = & \left. \mu \eta_{i_{6} j_{6}} \cal H \mit
_{(i_{6})} \rm \hat{\Phi}_{(j_{6})} \right|_{\theta \theta} +
                                    \left. g_{e I} \eta_{i_{6} j_{6}} \cal H
\mit _{(i_{6})} \pounds_{(j_{6})} \cal R \mit ^{I} \right|_{\theta \theta} +\\
                              &   & \left. g_{d I}\eta_{i_{6} j_{6}} \cal H
\mit _{(i_{6})} \cal Q \mit _{(j_{6}) \alpha} \cal D \mit ^{I}_{\alpha}
\right|_{\theta \theta} +
                                    \left. g_{u I}\eta_{i_{6} j_{6}}
\hat{\Phi}_{(j_{6})} \cal Q \mit _{(i_{6}) \alpha} \cal U \mit ^{I}_{\alpha}
\right|_{\theta \theta} +
                                    \rm h.c.
\end{array}
\label{eq2c*p*2-12}
\end{equation}
\begin{equation}
\begin{array}{lll}
\cal L \mit _{soft-breaking}& = & - m^{2}_{H} H^{\ast}_{(i_{6})} H_{(i_{6})} -
m^{2}_{\phi} \phi^{\ast}_{(i_{6})} \phi_{(i_{6})} \\
                            &   & - (m^{2}_{L}) L^{\prime \ast}_{(i_{6})}
L^{\prime}_{(i_{6})} - (m^{2}_{R})^{I J} R^{\prime I \ast} R^{\prime J} \\
                            &   & - m^{2}_{Q} Q^{\prime \ast}_{(i_{6}) \alpha}
Q^{\prime}_{(i_{6}) \alpha}
                                - (m^{2}_{D})^{I J} D^{\prime I \ast}_{\alpha}
D^{\prime J}_{\alpha} - (m^{2}_{U})^{I J} U^{\prime I \ast}_{\alpha} U^{\prime
J}_{\alpha} \\
                            &   & + m_{1} \left[  \lambda^{a}_{G}
\lambda^{a}_{G} + \rm h.c. \mit \right] + m_{2} \left[  \lambda^{i}_{A}
\lambda^{i}_{A} + \rm h.c. \mit \right]
                                  + m_{3} \left[  \lambda_{B} \lambda_{B} + \rm
h.c. \mit \right]  \\
                            &   & + \left[ h \eta_{i_{6} j_{6}} H_{(i_{6})}
\phi_{(j_{6})} + h_{e I} \eta_{i_{6} j_{6}} H_{(i_{6})}
                                  L^{\prime}_{(j_{6})} R^{\prime I}\right. \\
                            &   &  \left.\;\;\;\;+ h_{d I}\eta_{i_{6} j_{6}}
H_{(i_{6})} Q^{\prime I}_{(j_{6}) \alpha} D^{\prime I}_{\alpha}
                                          + h_{u I} \eta_{i_{6} j_{6}}
\phi_{(i_{6})} Q^{\prime}_{(i_{6}) \alpha} U^{\prime I}_{\alpha} + \rm
h.c.\right]
 \end{array}
\label{eq2d*p*2-12}
\end{equation}
The derivation of the preceding lagrangian was done in analogy to that of the
minimal supersymmetric standard model (MSSM) \cite{love}, \cite{Rosiek}.
Some relevant formulas are found in the appendix . In equation
\ref{eq2a*p*2-12},
\begin{equation}
k_{3} \ni \rm Tr \mit \left( Y^{a} Y^{b} \right) = k_{3} \delta^{a
b}\,,\;\;\;\; k_{3} > \rm 0
\end{equation}
and
\begin{equation}
k \ni \rm Tr \mit \left( T^{(i)} T^{(j)} \right) = k \delta^{i j}\,,\;\;\;\; k
> \rm 0.
\end{equation}

To be able to do calculations from the  preceding lagrangian, we have to deal
with the component fields of the superfields. Instead of writing down
the expansions for all the superfield expressions in equations
\ref{eq2a*p*2-12} to \ref{eq2c*p*2-12}, we will just show the expansion for
``prototype'' structures and then other terms which have similar structures are
calculated by substituting analogous quantities. These prototype structures
are derived using equations \ref{eq5*p*ApA-2} to \ref{eq7*p*ApA-2} of the
appendix , the following prototype structures
($V$, $\Phi_{i}$ are vector and chiral superfields respectively)
\begin{equation}
V(x) = - \theta \sigma^{\mu} \bar{\theta} v_{\mu} + i \theta \theta
\bar{\theta} \bar{\lambda} - i \bar{\theta} \bar{\theta} \theta \lambda
      + \frac{1}{2} \theta \theta \bar{\theta} \bar{\theta} D
\end{equation}
\begin{equation}
\begin{array}{ccl}
\Phi(x) & = & A(y) + \sqrt{2} \theta \psi(y) + \theta \theta F(y) \\
                 & = & A(x) + i \theta \sigma^{\mu} \bar{\theta} \partial_{\mu}
A(x) + \frac{1}{4} \theta \theta \bar{\theta} \bar{\theta}
                       \Box A(x) +\\
                 &   & \sqrt{2} \theta \psi(x) - \frac{i}{\sqrt{2}} \theta
\theta \partial_{\mu} \psi(x) \sigma^{\mu} \bar{\theta}
                       + \theta \theta F(x)
\end{array}
\end{equation}
and the definitions,
\begin{equation}
D_{\eta} \equiv \frac{\partial}{\partial \theta^{\eta}} + i \sigma^{\mu}_{\eta
\dot{\eta}} \bar{\theta}^{\dot{\eta}} \partial_{\mu}
\end{equation}
\begin{equation}
\bar{D}_{\dot{\eta}} \equiv -\frac{\partial}{\partial
\bar{\theta}^{\dot{\eta}}} - i \theta^{\eta} \sigma^{\mu}_{\eta \dot{\eta}}
\partial_{\mu}. \end{equation}

One has to also take note of the fermionic, bosonic and auxiliary field
components of the superfields as given in tables
\ref{tbl1*p*2-7}, \ref{tbl1*p*2-8}, \ref{tbl1*p*2-9}, \ref{tbl1*p*2-10} and
\ref{tbl2*p*2-10}. The auxiliary fields
$D^{a}_{G}$,..., etc. and $F_{l\,(i_{6})}$,..., etc. (see tables
\ref{tbl1*p*2-7} and \ref{tbl2*p*2-10}) are eliminated from the
lagrangian through the equations of motion,
\begin{equation}
\frac{\partial \cal L}{\partial \phi} - \partial_{\mu} \left( \frac{\partial
\cal L}{\partial \left( \partial_{\mu} \phi \right)} \right) = 0
\end{equation}
for a field $\phi$. Note that the fundamental equations given above and their
manipulation were taken from reference
\cite{Wess-Bagger}. Since their metric tensor is different from what we use
here, we have to adjust the signs of the terms
involving the component fields.

In the following formulas, one will notice terms like $\left[ \frac{1}{2} \cdot
\frac{1}{3} \right]$ or $(\frac{1}{2}) (\frac{1}{3})$. We purposely did not
simplify this to emphasize the fact that we multiply $\frac{1}{2}$ by the
hypercharge
of the multiplet. For example, in equations \ref{eq3*p*2-14}, \ref{eq1a*p*2-15}
and \ref{eq1b*p*2-15} below, we have
$(\frac{1}{2}) (\frac{1}{3})$ and $\left[ \frac{1}{2} \cdot \frac{1}{3}
\right]$ because the $\cal Q \mit _{(i_{6})}$
multiplet has a hypercharge $\frac{1}{3}$ as in table~\ref{tbl1*p*2-11}.

For the $\cal L \mit _{YM}$, we use the prototype structure
\begin{equation}
\begin{array}{ccl}
\frac{1}{4 k_{3} (2 g_{3})^{2}} \rm Tr \mit \left[  \left.W^{\eta}_{G} W_{G
\eta} \right|_{\theta \theta} +
                                                     \left. \rm \bar{\mit W}_{G
\dot{\eta}} \bar{\mit W}^{\dot{\eta}}_{G} \right|_{\rm \bar{\theta}
\bar{\theta}} \right]
& = & - \frac{1}{4} G^{\mu \nu}_{a} G_{\mu \nu a} + i \rm \bar{\lambda}_{\mit G
a} \bar{\sigma}^{\mu}\left[ \cal D \mit _{\mu} \lambda_{G} \right]_{a} \\
  &  & + \frac{1}{2} D_{G a} D_{G a}
\end{array}
\end{equation}
where
\begin{equation}
G^{\mu \nu}_{a} \equiv \partial^{\mu} G^{\nu}_{a} - \partial^{\nu} G^{\mu}_{a}
- g_{3} f_{abc} G^{\mu}_{b} G^{\nu}_{c}
\end{equation}
\begin{equation}
\left[ \cal D \mit _{\mu} \lambda_{G} \right]_{a} = \partial_{\mu} \lambda_{G
a} - g_{3} f_{a b c} G_{\mu b} \lambda_{G c}\;.
\end{equation}

For the $\cal L \mit _{kin}$, we use the prototype structure
\begin{equation}
\begin{array}{l}
\left. \cal Q \mit ^{\dagger}_{(i_{6}) \alpha}
                         e^{2 \left[g_{3} Y^{a}_{\alpha \beta} \cal G \mit _{a}
\delta_{i_{6} j_{6}} +
                                    g_{sp} \delta_{\alpha \beta}
T^{(i)}_{i_{6} j_{6}} \cal A \mit _{i} +
                                    g_{1} \left( \frac{1}{2}\right) \left(
\frac{1}{3} \right) \delta_{\alpha \beta} \delta_{i_{6} j_{6}} \rm \hat{\cal B}
                             \right]}
                         \cal Q \mit _{(j_{6}) \beta} \right|_{\theta \theta
\rm \bar{\theta} \bar{\theta}} = \\
\nabla_{\mu} Q^{\prime \ast}_{(i_{6}) \gamma} \nabla^{\mu} Q^{\prime}_{(i_{6})
\gamma} + i \bar{\psi}^{\prime}_{q\, (i_{6}) \alpha} \bar{\sigma}^{\mu}
\nabla_{\mu}
        \psi^{\prime}_{q\, (i_{6}) \alpha} + F^{\ast}_{q\, (i_{6}) \alpha}
F_{q\, (i_{6}) \alpha} \\
+ i \sqrt{2} g_{3} Y^{a}_{\alpha \beta} \left[ Q^{\prime \ast}_{(i_{6}) \alpha}
\psi^{\prime}_{q\, (i_{6}) \beta} \lambda_{G a}
                                               - \bar{\lambda}_{G a}
\bar{\psi}^{\prime}_{q\, (i_{6}) \alpha} Q^{\prime}_{(i_{6}) \beta}\right] \\
+ i \sqrt{2} g_{sp} T^{(i)}_{i_{6} j_{6}} \left[ Q^{\prime \ast}_{(i_{6})
\alpha} \psi^{\prime}_{q\, (j_{6}) \alpha} \lambda_{A i}
                                               - \bar{\lambda}_{A i}
Q^{\prime}_{(j_{6}) \alpha} \bar{\psi}^{\prime}_{q\, (i_{6}) \alpha} \right] \\
+ i \sqrt{2} g_{1} \left[ \frac{1}{2} \cdot \frac{1}{3} \right] \left[
Q^{\prime \ast}_{(i_{6}) \alpha} \lambda_{B} \psi^{\prime}_{q\, (i_{6}) \alpha}
                                               - Q^{\prime}_{(i_{6}) \alpha}
\bar{\lambda}_{B}  \bar{\psi}^{\prime}_{q\, (i_{6}) \alpha} \right] \\
+ g_{3} D_{G a} Q^{\prime \ast}_{(i_{6}) \alpha} Y^{a}_{\alpha \beta}
Q^{\prime}_{(i_{6}) \beta} + g_{sp} D_{A i} Q^{\prime \ast}_{(i_{6}) \alpha}
T^{(i)}_{i_{6} j_{6}}
  Q^{\prime}_{(j_{6}) \alpha}\\
 + g_{1} D_{B} Q^{\prime \ast}_{(i_{6}) \alpha} \left[ \frac{1}{2} \cdot
\frac{1}{3} \right] Q^{\prime}_{(i_{6}) \alpha}
\end{array}
\label{eq3*p*2-14}
\end{equation}
where
\begin{equation}
\begin{array}{lll}
\nabla^{\mu} Q^{\prime}_{(i_{6}) \gamma} & \equiv & \partial^{\mu}
Q^{\prime}_{(i_{6}) \gamma} + i g_{3} G^{\mu}_{b} Y^{b}_{ \gamma \beta}
Q^{\prime}_{(i_{6}) \beta}
                                                  + i  g_{sp} A^{\mu}_{j}
T^{(j)}_{i_{6} j_{6}} Q^{\prime}_{(j_{6}) \gamma} \\
                                            &       & + i g_{1} B^{\mu} \left[
\frac{1}{2} \cdot \frac{1}{3} \right] Q^{\prime}_{(j_{6}) \gamma}
\end{array}
\label{eq1a*p*2-15}
\end{equation}
\begin{equation}
\begin{array}{lll}
\nabla^{\mu} \psi^{\prime}_{q\, (i_{6}) \alpha} & \equiv & \partial^{\mu}
\psi^{\prime}_{q\, (i_{6}) \alpha} + i g_{3} G^{\mu}_{a} Y^{a}_{ \alpha \beta}
\psi^{\prime}_{q\, (i_{6}) \beta}
                                                  + i  g_{sp} A_{\mu j}
T^{(j)}_{i_{6} j_{6}} \psi^{\prime}_{q\, (j_{6}) \alpha} \\
                                            &       & + i g_{1} B^{\mu} \left[
\frac{1}{2} \cdot \frac{1}{3} \right] \psi^{\prime}_{q\, (i_{6}) \alpha}\;.
\end{array}
\label{eq1b*p*2-15}
\end{equation}

For the  $\cal L \mit _{superpotential}$ we have the prototype structure
\begin{equation}
\left. \mu \eta_{i_{6} j_{6}} \cal H \mit _{(i_{6})} \rm \hat{\Phi}_{(j_{6})}
\right|_{\theta \theta} =
\mu \eta_{i_{6} j_{6}} \left[ H_{(i_{6})}  F_{\phi (j_{6})} + \phi_{(j_{6})}
F_{H (i_{6})} - \psi_{H (i_{6})} \psi_{\phi (j_{6})}\right]
\end{equation}
\begin{equation}
\begin{array}{lll}
\left. g_{e I} \eta_{i_{6} j_{6}} \cal H \mit _{(i_{6})} \pounds_{(j_{6})} \cal
R \mit ^{I} \right|_{\theta \theta} & = &
g_{e I} \eta_{i_{6} j_{6}}\left[ H_{(i_{6})} L^{\prime}_{(j_{6})} F^{I}_{R} +
H_{(i_{6})} F_{l\, (j_{6})} R^{\prime I} +
F_{H\, (i_{6})}  L^{\prime}_{(j_{6})} R^{\prime I} \right. \\
 &  & \left. - H_{(i_{6})} \psi^{\prime}_{l \, (j_{6})} \psi^{\prime I}_{R} -
L^{\prime}_{(j_{6})} \psi_{H (i_{6})} \psi^{\prime I}_{R}
- \psi_{H (i_{6})} \psi^{\prime}_{l \, (j_{6})} R^{\prime I} \right]
\end{array}
\end{equation}

\clearpage
\section{$B^{0}_{d} \bar{B}^{0}_{d}$ mixing in SUSY $Sp(6)$}
\label{bbmix}

\input FEYNMAN
\input psfig
We now study $B^{0}_{d} \bar{B}^{0}_{d}$ mixing in the framework of SUSY
$Sp(6)$. In a previous paper \cite{Blad-Kuo}, we
investigated $B^{0}_{d} \bar{B}^{0}_{d}$ in the framework of the ordinary
(non-SUSY) $Sp(6)$. It was shown that the tree-level
contributions due to the extra $Z^{\prime}$ of the $Sp(6)$ model enhances the
FCNC contribution to the mixing parameter
$x_{d}$. Along the same lines, the gluino contributions of the MSSM, as
described in reference \cite{Franzini},  also tend to
enhance $x_{d}$.
 These, however,  proved useful for lower values of the top mass $m_{t}$. With
the recent experimental limits
\cite{abe} on $m_{t}$ of $158 \stackrel{<}{\sim} m_{t} \stackrel{<}{\sim} 194$
Gev, the SM contribution to $B^{0}_{d} \bar{B}^{0}_{d}$ mixing
maybe high enough to be within the ARGUS, CLEO result \cite{Schroder} of $0.57
\stackrel{<}{\sim} x_{d} \stackrel{<}{\sim} 0.77$.
This implies that models which enhance $x_{d}$ may not be so appealing. Because
of the uncertain mass of the $Z^{\prime}$, we can
always find a value for it to fit the value for $x_{d}$. However, with the high
$m_{t}$, the $m_{Z^{\prime}}$ becomes uninterestingly
large. It turns out that in the SUSY $Sp(6)$ model, the additional
contributions of the $\tilde{z}^{\prime}$ ($Z^{\prime}$-ino) tend
to cancel the other contributions. This leads to a lower $m_{Z^{\prime}}$
value. It should be stated, however, that the calculation of
$x_{d}$ involve uncertain parameters that make the conclusions based on the
numerics less definitive than we would want to. Nevertheless,
the relative numerical relationships can give us definite statements with
respect to possible effects.

The SM and $Sp(6)$ model contributions had been discussed in reference
\cite{Blad-Kuo}. Here, we will focus our discussion on the SUSY
contributions. Let us first discuss the analysis of the contributions of the
MSSM \footnote{For an excellent paper in which the notation
is close to ours on MSSM, see reference \cite{Rosiek}.}
 to $B^{0}_{d} \bar{B}^{0}_{d}$ mixing.

Because of the richer particle spectrum of a SUSY theory, we expect new
contributions to FCNC. Of course, these FCNC contributions must
not be too additively large for SUSY to remain acceptable.

The three additional contributions to $B^{0}_{d} \bar{B}^{0}_{d}$ mixing  at
the one loop level in the MSSM are due
\begin{enumerate}
   \item to the physical charged scalar higgs;
   \item to the SUSY partner of the $W$ (or more precisely the physical
charginos) and the charged scalar higgs and
   \item to the neutralinos and gluinos.
\end{enumerate}

For the first contribution in item 2 above, we basically replace the $W$ by its
SUSY partner, $\tilde{w}$ and the $u$, $c$ and $t$
by their SUSY partners $\tilde{u}$, $\tilde{c}$ and $\tilde{t}$ respectively.
Item 3 above on the other hand, is less obvious. It was only
realized in 1983 \cite{Duncan} and is unique to SUSY theories since it has no
SM analogue. Let us briefly explain why item 3 above is
less obvious.

Working in component fields, upon the elimination of the auxiliary fields from
the lagrangian of section \ref{susy-sp1} and with the
appropriate $\cal L \mit _{soft-breaking}$ terms of equation \ref{eq2d*p*2-12},
we can calculate the squark mass matrix (see page
3468 of reference \cite{Rosiek}). Let us assume for the moment that the mass
matrix $(m^{2}_{Q})^{I J}$ of page 3468 reference
\cite{Rosiek} is diagonal, i.e. $(m^{2}_{Q})^{I J} \longrightarrow
(m^{2}_{Q})^{I J} \delta^{I J}$ (usually this is assumed,
see for example equation 2.2 of reference \cite{Gabbiani}). We can then write
the squared mass matrix of the superpartner
of the left-handed down squarks as
\begin{equation}
M^{2}_{Q_{2}} = \mu^{(0)}_{Q_{2}} {\bf 1} + \mu^{(1)}_{Q_{2}}
M^{\dagger}_{q_{d}} M_{q_{d}}
\label{eq1*p*3-2}
\end{equation}
where $M_{q_{d}}$ is the mass matrix of the down quarks and $\mu^{(0)}_{Q_{2}}$
and $\mu^{(1)}_{Q_{2}}$ are some parameter
coefficients. Hence, in equation \ref{eq1*p*3-2}, we see that diagonalizing the
down quark matrix through a redefinition of
fields will automatically diagonalize $M^{2}_{Q_{2}}$. Thus a coupling
\begin{equation}
\label{eq1*p*3-3}
\Lambda^{a}_{G} \bar{q}^{\prime I}_{d} Q^{\prime J}_{(2)}
\end{equation}
where $q^{\prime I}_{d}$ and $Q^{\prime J}_{(2)}$ are the initial down quark
and squark fields and $\Lambda^{a}_{G}$
is the gluino, will be \underline{diagonalized} upon the redefinition of the
quark and squark fields since the unitary matrices
which are used to redefine them to  get their mass eigenstates are the same.
Hence, no FCNC occurs. However, when one renormalizes
$M^{2}_{Q_{2}}$ from its initial value  in equation \ref{eq1*p*3-2} at the
superlarge scale down to the $m_{W}$ scale, it is shown
\cite{Duncan} that due to the term $\left. \epsilon_{i_{2} j_{2}} u^{I J} \cal
H \mit ^{2}_{(i_{2})} \cal Q \mit ^{I}_{(j_{2}) \alpha} \cal U \mit
^{J}_{\alpha} \right|_{\theta \theta}$ in the MSSM superpotential (see for
example section 5.2.2 reference \cite{blado}),
 there arises a term proportional to the square of the mass matrix
($M^{\dagger}_{q_{u}} M_{q_{u}}$) of the up-type quarks. Equation
\ref{eq1*p*3-2} becomes
\begin{equation}
M^{2}_{Q_{2}} = \mu^{(0)}_{Q_{2}} {\bf 1} + \mu^{(1)}_{Q_{2}}
M^{\dagger}_{q_{d}} M_{q_{d}} + \mu^{(2)}_{Q_{2}} M^{\dagger}_{q_{u}} M_{q_{u}}
\end{equation}
Hence, diagonalizing $M^{\dagger}_{q_{d}} M_{q_{d}}$ \underline{does not}
diagonalize $M^{2}_{Q_{2}}$. This essentially means that the unitary
matrices used to redefine the quark fields to get their mass eigenstates will
in general be \underline{different} from the unitary matrices
used to redefine the squark fields. Hence, the coupling in equation
\ref{eq1*p*3-3} will \underline{not be diagonalized} leading to FCNC.
The coefficient $\mu^{(2)}_{Q_{2}}$ is calculated by solving the set of
renormalization group equations for the
evolution of the SUSY quantities.

It turns out that the contribution due to the gluinos is the most dominant MSSM
contribution to $B^{0}_{d} \bar{B}^{0}_{d}$ mixing due
to the strong coupling $\alpha_{s}$ ($= \frac{g^{2}_{3}}{4 \pi}$).

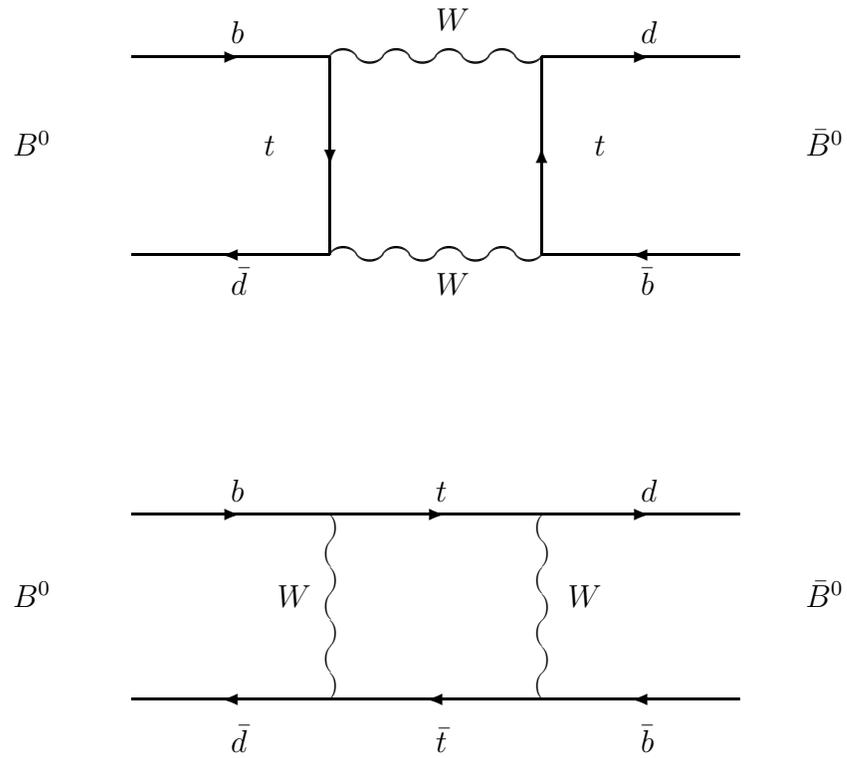
\begin{figure}[htbp]
\caption{Dominant SM contributions to $B^{0}_{d} \bar{B}^{0}_{d}$ mixing}
\begin{center}
\normalsize
\begin{picture}(10000,10000)(2000,0)
%
\bigphotons
\THICKLINES
\drawline\photon[\E\REG](5000,5000)[8]
\put(\pmidx,6000){$W$}
%
\drawline\fermion[\W\REG](\particlefrontx,\particlefronty)[7500]
\drawarrow[\E\ATBASE](\pmidx,\pmidy)
\put(\pmidx,5500){$b$}
%
\drawline\fermion[\S\REG](\particlefrontx,\particlefronty)[7500]
\drawarrow[\S\ATBASE](\pmidx,\pmidy)
\put(2500,\pmidy){$t$}
\put(-7000,\pmidy){$B^0$}
%
\drawline\fermion[\W\REG](\particlebackx,\particlebacky)[7500]
\drawarrow[\W\ATBASE](\pmidx,\pmidy)
\put(\pmidx,-4000){$\bar{d}$}
%
\drawline\photon[\E\REG](\particlefrontx,\particlefronty)[8]
\put(\pmidx,-4000){$W$}
%
\drawline\fermion[\E\REG](\particlebackx,\particlebacky)[7500]
\drawarrow[\W\ATBASE](\pmidx,\pmidy)
\put(\pmidx,-4000){$\bar{b}$}
%
\drawline\fermion[\N\REG](\particlefrontx,\particlefronty)[7500]
\drawarrow[\N\ATBASE](\pmidx,\pmidy)
\put(15000,\pmidy){$t$}
\put(23000,\pmidy){$\bar{B}^0$}
%
\drawline\fermion[\E\REG](\fermionbackx,\fermionbacky)[7500]
\drawarrow[\E\ATBASE](\pmidx,\pmidy)
\put(\pmidx,5500){$d$}
\end{picture}
\vskip 1.0in
\begin{picture}(10000,10000)(2000,0)
\THICKLINES
%
\drawline\fermion[\E\REG](5000,5000)[\photonlengthx]
\drawarrow[\E\ATBASE](\pmidx,\pmidy)
\put(\pmidx,5500){$t$}
%
\drawline\fermion[\W\REG](\particlefrontx,\particlefronty)[7500]
\drawarrow[\E\ATBASE](\pmidx,\pmidy)
\put(\pmidx,5500){$b$}
%
\drawline\photon[\S\REG](\particlefrontx,\particlefronty)[7]
\put(3000,\pmidy){$W$}
\put(-7000,\pmidy){$B^0$}
%
\drawline\fermion[\W\REG](\particlebackx,\particlebacky)[7500]
\drawarrow[\W\ATBASE](\pmidx,\pmidy)
\put(\pmidx,-4000){$\bar{d}$}
%
\startphantom
\drawline\photon[\E\REG](5000,5000)[8]
\stopphantom
%
\drawline\fermion[\E\REG](\fermionfrontx,\fermionfronty)[\photonlengthx]
\drawarrow[\W\ATBASE](\pmidx,\pmidy)
\put(\pmidx,-4000){$\bar{t}$}
%
\drawline\fermion[\E\REG](\particlebackx,\particlebacky)[7500]
\drawarrow[\W\ATBASE](\pmidx,\pmidy)
\put(\pmidx,-4000){$\bar{b}$}
%
\drawline\photon[\N\REG](\particlefrontx,\particlefronty)[7]
\put(14000,\pmidy){$W$}
\put(23000,\pmidy){$\bar{B}^0$}
%
\drawline\fermion[\E\REG](\particlebackx,\particlebacky)[7500]
\drawarrow[\E\ATBASE](\pmidx,\pmidy)
\put(\pmidx,5500){$d$}
\end{picture}
\end{center}
\label{fig1*p*3-4}
\end{figure}

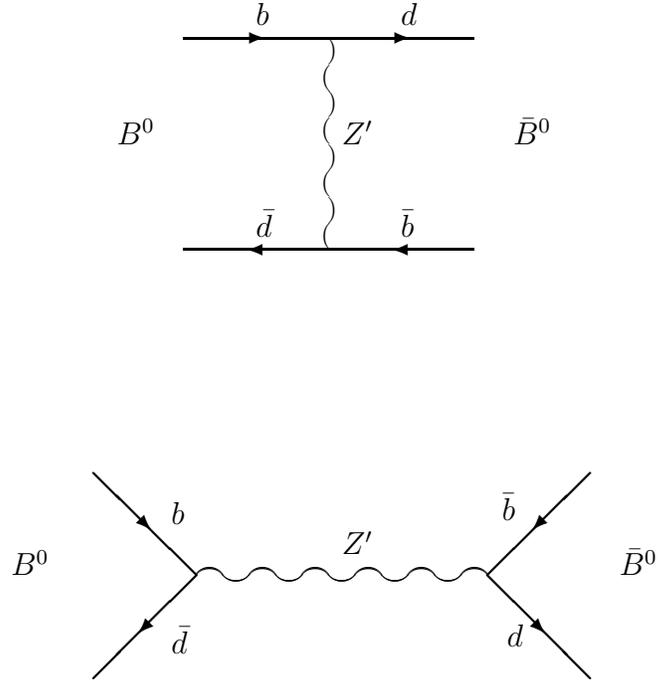
\begin{figure}[htbp]
\caption{$Sp(6)$ model tree level contributions to $B^{0}_{d} \bar{B}^{0}_{d}$
mixing due to the extra $Z^{\prime}$}
\normalsize
\begin{center}
\begin{picture}(10000,10000)(1000,0)
\bigphotons
\drawline\photon[\S\REG](5000,8000)[8]
\put(5500,\pmidy){$Z^{\prime}$}
\put(-3000,\pmidy){$B^{0}$}
\put(12000,\pmidy){$\bar{B}^{0}$}
\THICKLINES
\drawline\fermion[\W\REG](\photonfrontx,\photonfronty)[5500]
\drawarrow[\E\ATBASE](\pmidx,\pmidy)
\put(\pmidx,8500){$b$}
\drawline\fermion[\E\REG](\photonfrontx,\photonfronty)[5500]
\drawarrow[\E\ATBASE](\pmidx,\pmidy)
\put(\pmidx,8500){$d$}
\drawline\fermion[\W\REG](\photonbackx,\photonbacky)[5500]
\drawarrow[\W\ATBASE](\pmidx,\pmidy)
\put(\pmidx,500){$\bar{d}$}
\drawline\fermion[\E\REG](\photonbackx,\photonbacky)[5500]
\drawarrow[\W\ATBASE](\pmidx,\pmidy)
\put(\pmidx,500){$\bar{b}$}
\end{picture}
\vskip 1.0in
\begin{picture}(10000,10000)
\bigphotons
\THICKLINES
\drawline\photon[\E\REG](-1000,5000)[11]
\put(-8000,5000){$B^{0}$}
\global\advance \photonbackx by 5000
\put(\photonbackx,5000){$\bar{B}^{0}$}
\global\advance \photonbackx by -5000
\global\advance \pmidx by 0
\put(\pmidx,5800){$Z^{\prime}$}
\drawline\fermion[\NW\REG](\photonfrontx,\photonfronty)[5500]
\drawarrow[\SE\ATBASE](\pmidx,\pmidy)
\global\advance \pmidx by 1000
\put(\pmidx,\pmidy){$b$}
\drawline\fermion[\SW\REG](\photonfrontx,\photonfronty)[5500]
\drawarrow[\SW\ATBASE](\pmidx,\pmidy)
\global\advance \pmidx by 1000
\global\advance \pmidy by -1000
\put(\pmidx,\pmidy){$\bar{d}$}
\drawline\fermion[\NE\REG](\photonbackx,\photonbacky)[5500]
\drawarrow[\SW\ATBASE](\pmidx,\pmidy)
\global\advance \pmidx by -1400
\put(\pmidx,7100){$\bar{b}$}
\drawline\fermion[\SE\REG](\photonbackx,\photonbacky)[5500]
\drawarrow[\SE\ATBASE](\pmidx,\pmidy)
\global\advance \pmidx by -1200
\put(\pmidx,2300){$d$}
\end{picture}
\end{center}
\label{fig2*p*3-4}
\end{figure}

We next discuss the contributions of SUSY $Sp(6)$ to $B^{0}_{d}
\bar{B}^{0}_{d}$ mixing. For completeness we point out the complete set
of (dominant) graphs in SUSY $Sp(6)$ which will contribute to $B^{0}_{d}
\bar{B}^{0}_{d}$ mixing in figures \ref{fig1*p*3-4},
\ref{fig2*p*3-4}, \ref{fig3*p*3-4} and \ref{fig1*p*3-5}.
%
%
%
\begin{figure}[htbp]
\caption{Dominant minimal supersymmetric standard model contributions to
$B^{0}_{d} \bar{B}^{0}_{d}$ mixing }
\normalsize
\begin{flushleft}
\begin{picture}(10000,10000)
%
\thicklines
\drawline\fermion[\E\REG](7000,5000)[5500]
\global\advance\pmidx by -1000
\put(\pmidx,6000){$\Lambda^{a}_{G}$}
%
\drawline\fermion[\W\REG](\particlefrontx,\particlefronty)[5500]
\drawarrow[\E\ATBASE](\pmidx,\pmidy)
\put(\pmidx,5500){$b$}
%
\drawline\scalar[\S\REG](\particlefrontx,\particlefronty)[3]
\put(3000,\pmidy){$Q^{I}_{(2)}$}
%
\drawline\fermion[\W\REG](\particlebackx,\particlebacky)[5500]
\drawarrow[\W\ATBASE](\pmidx,\pmidy)
\global\advance\pmidy by -2000
\put(\pmidx,\pmidy){$\bar{d}$}
%
\drawline\fermion[\E\REG](\particlefrontx,\particlefronty)[5500]
\global\advance\pmidy by -2000
\global\advance\pmidx by -1000
\put(\pmidx,\pmidy){$\Lambda^{b}_{G}$}
%
\drawline\fermion[\E\REG](\particlebackx,\particlebacky)[5500]
\drawarrow[\E\ATBASE](\pmidx,\pmidy)
\global\advance\pmidy by -2000
\put(\pmidx,\pmidy){$d$}
%
\drawline\scalar[\N\REG](\particlefrontx,\particlefronty)[3]
\global\advance\pmidx by 1000
\put(\pmidx,\pmidy){$Q^{J}_{(2)}$}
%
\drawline\fermion[\E\REG](\scalarbackx,\scalarbacky)[5500]
\drawarrow[\W\ATBASE](\pmidx,\pmidy)
\put(\pmidx,5500){$\bar{b}$}
%
\thicklines
\drawline\fermion[\E\REG](30000,5000)[5500]
\global\advance\pmidx by -1000
\put(\pmidx,6000){$\Lambda^{a}_{G}$}
%
\drawline\fermion[\W\REG](\particlefrontx,\particlefronty)[5500]
\drawarrow[\E\ATBASE](\pmidx,\pmidy)
\put(\pmidx,5500){$b$}
%
\drawline\scalar[\S\REG](\particlefrontx,\particlefronty)[3]
\put(26000,\pmidy){$Q^{I}_{(2)}$}
%
\drawline\fermion[\W\REG](\particlebackx,\particlebacky)[5500]
\drawarrow[\W\ATBASE](\pmidx,\pmidy)
\global\advance\pmidy by -2000
\put(\pmidx,\pmidy){$\bar{d}$}
%
\drawline\fermion[\E\REG](\particlefrontx,\particlefronty)[5500]
\global\advance\pmidy by -2000
\global\advance\pmidx by -1000
\put(\pmidx,\pmidy){$\Lambda^{b}_{G}$}
%
\drawline\fermion[\E\REG](\particlebackx,\particlebacky)[5500]
\drawarrow[\W\ATBASE](\pmidx,\pmidy)
\global\advance\pmidy by -2000
\put(\pmidx,\pmidy){$\bar{b}$}
%
\drawline\scalar[\N\REG](\particlefrontx,\particlefronty)[3]
\global\advance\pmidx by 1000
\put(\pmidx,\pmidy){$Q^{J}_{(2)}$}
%
\drawline\fermion[\E\REG](\scalarbackx,\scalarbacky)[5500]
\drawarrow[\E\ATBASE](\pmidx,\pmidy)
\put(\pmidx,5500){$d$}
\end{picture}
\vskip 0.5in
\begin{picture}(10000,10000)
%
\thicklines
\drawline\scalar[\E\REG](7000,5000)[3]
\global\advance\pmidx by -1000
\put(\pmidx,6000){$Q^{I}_{(2)}$}
%
\drawline\fermion[\W\REG](\particlefrontx,\particlefronty)[5500]
\drawarrow[\E\ATBASE](\pmidx,\pmidy)
\put(\pmidx,5500){$b$}
\startphantom
\drawline\scalar[\S\REG](\particlefrontx,\particlefronty)[3]
\stopphantom
%
\drawline\fermion[\S\REG](\particlefrontx,\particlefronty)[\scalarlengthy]
\put(4500,\pmidy){$\Lambda^{a}_{G}$}
%
\drawline\fermion[\W\REG](\particlebackx,\particlebacky)[5500]
\drawarrow[\W\ATBASE](\pmidx,\pmidy)
\global\advance\pmidy by -2000
\put(\pmidx,\pmidy){$\bar{d}$}
%
\drawline\scalar[\E\REG](\particlefrontx,\particlefronty)[3]
\global\advance\pmidy by -2000
\global\advance\pmidx by -1000
\put(\pmidx,\pmidy){$Q^{J}_{(2)}$}
%
\drawline\fermion[\E\REG](\particlebackx,\particlebacky)[5500]
\drawarrow[\W\ATBASE](\pmidx,\pmidy)
\global\advance\pmidy by -2000
\put(\pmidx,\pmidy){$\bar{b}$}
\startphantom
\drawline\scalar[\S\REG](\particlefrontx,\particlefronty)[3]
\stopphantom
%
\drawline\fermion[\N\REG](\particlefrontx,\particlefronty)[\scalarlengthy]
\global\advance\pmidx by 1000
\put(\pmidx,\pmidy){$\Lambda^{b}_{G}$}
%
\drawline\fermion[\E\REG](\fermionbackx,\fermionbacky)[5500]
\drawarrow[\E\ATBASE](\pmidx,\pmidy)
\put(\pmidx,5500){$d$}
%
\thicklines
\drawline\fermion[\E\REG](30000,5000)[6000]
\global\advance\pmidx by -1000
\put(\pmidx,6000){$\Lambda^{a}_{G}$}
%
\drawline\fermion[\W\REG](\particlefrontx,\particlefronty)[5500]
\drawarrow[\E\ATBASE](\pmidx,\pmidy)
\put(\pmidx,5500){$b$}
%
\drawline\scalar[\SE\REG](\particlefrontx,\particlefronty)[4]
\global\advance\pmidx by 2000
\global\advance\pmidy by -1000
\put(\pmidx,\pmidy){$Q^{I}_{(2)}$}
%
\drawline\fermion[\E\REG](\particlebackx,\particlebacky)[5500]
\drawarrow[\E\ATBASE](\pmidx,\pmidy)
\global\advance\pmidy by -2000
\put(\pmidx,\pmidy){$d$}
%
\drawline\fermion[\W\REG](\particlefrontx,\particlefronty)[6000]
\global\advance\pmidy by -2000
\global\advance\pmidx by -1000
\put(\pmidx,\pmidy){$\Lambda^{b}_{G}$}
%
\drawline\fermion[\W\REG](\particlebackx,\particlebacky)[5500]
\drawarrow[\W\ATBASE](\pmidx,\pmidy)
\global\advance\pmidy by -2000
\put(\pmidx,\pmidy){$\bar{d}$}
%
\drawline\scalar[\NE\REG](\particlefrontx,\particlefronty)[4]
\global\advance\pmidx by -4000
\global\advance\pmidy by -1000
\put(\pmidx,\pmidy){$Q^{J}_{(2)}$}
%
\drawline\fermion[\E\REG](\scalarbackx,\scalarbacky)[5500]
\drawarrow[\W\ATBASE](\pmidx,\pmidy)
\put(\pmidx,5500){$\bar{b}$}
\end{picture}
\end{flushleft}
\label{fig3*p*3-4}
\end{figure}
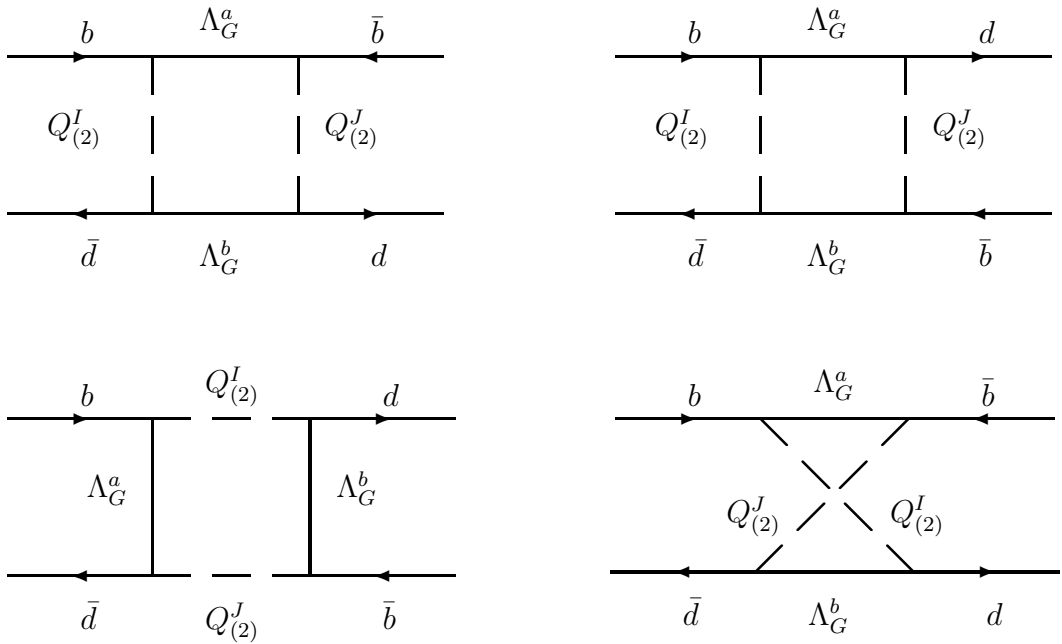
%
\begin{figure}[htbp]
\caption{Dominant SUSY $Sp(6)$ contributions to $B^{0}_{d} \bar{B}^{0}_{d}$
mixing due to the $\tilde{z}^{\prime}$ ($Z^{\prime}$-ino)}
\normalsize
\begin{flushleft}
\begin{picture}(10000,10000)(0, -2000)
%
\thicklines
\drawline\fermion[\E\REG](7000,5000)[5500]
\global\advance\pmidx by -1000
\put(\pmidx,6000){$\Lambda^{a}_{G}$}
%
\drawline\fermion[\W\REG](\particlefrontx,\particlefronty)[5500]
\drawarrow[\E\ATBASE](\pmidx,\pmidy)
\put(\pmidx,5500){$b$}
%
\drawline\scalar[\S\REG](\particlefrontx,\particlefronty)[3]
\put(3000,\pmidy){$Q^{I}_{(2)}$}
%
\drawline\fermion[\W\REG](\particlebackx,\particlebacky)[5500]
\drawarrow[\W\ATBASE](\pmidx,\pmidy)
\global\advance\pmidy by -2000
\put(\pmidx,\pmidy){$\bar{d}$}
%
\drawline\fermion[\E\REG](\particlefrontx,\particlefronty)[5500]
\global\advance\pmidy by -2000
\global\advance\pmidx by -1000
\put(\pmidx,\pmidy){$\tilde{z}^{\prime}$}
%
\drawline\fermion[\E\REG](\particlebackx,\particlebacky)[5500]
\drawarrow[\E\ATBASE](\pmidx,\pmidy)
\global\advance\pmidy by -2000
\put(\pmidx,\pmidy){$d$}
%
\drawline\scalar[\N\REG](\particlefrontx,\particlefronty)[3]
\global\advance\pmidx by 1000
\put(\pmidx,\pmidy){$Q^{J}_{(2)}$}
%
\drawline\fermion[\E\REG](\scalarbackx,\scalarbacky)[5500]
\drawarrow[\W\ATBASE](\pmidx,\pmidy)
\put(\pmidx,5500){$\bar{b}$}
%
\thicklines
\drawline\fermion[\E\REG](30000,5000)[5500]
\global\advance\pmidx by -1000
\put(\pmidx,6000){$\Lambda^{a}_{G}$}
%
\drawline\fermion[\W\REG](\particlefrontx,\particlefronty)[5500]
\drawarrow[\E\ATBASE](\pmidx,\pmidy)
\put(\pmidx,5500){$b$}
%
\drawline\scalar[\S\REG](\particlefrontx,\particlefronty)[3]
\put(26000,\pmidy){$Q^{I}_{(2)}$}
%
\drawline\fermion[\W\REG](\particlebackx,\particlebacky)[5500]
\drawarrow[\W\ATBASE](\pmidx,\pmidy)
\global\advance\pmidy by -2000
\put(\pmidx,\pmidy){$\bar{d}$}
%
\drawline\fermion[\E\REG](\particlefrontx,\particlefronty)[5500]
\global\advance\pmidy by -2000
\global\advance\pmidx by -1000
\put(\pmidx,\pmidy){$\tilde{z}^{\prime}$}
%
\drawline\fermion[\E\REG](\particlebackx,\particlebacky)[5500]
\drawarrow[\W\ATBASE](\pmidx,\pmidy)
\global\advance\pmidy by -2000
\put(\pmidx,\pmidy){$\bar{b}$}
%
\drawline\scalar[\N\REG](\particlefrontx,\particlefronty)[3]
\global\advance\pmidx by 1000
\put(\pmidx,\pmidy){$Q^{J}_{(2)}$}
%
\drawline\fermion[\E\REG](\scalarbackx,\scalarbacky)[5500]
\drawarrow[\E\ATBASE](\pmidx,\pmidy)
\put(\pmidx,5500){$d$}
\end{picture}
\vskip 0.5in
\begin{picture}(10000,10000)(0, -2000)
%
\thicklines
\drawline\scalar[\E\REG](7000,5000)[3]
\global\advance\pmidx by -1000
\put(\pmidx,6000){$Q^{I}_{(2)}$}
%
\drawline\fermion[\W\REG](\particlefrontx,\particlefronty)[5500]
\drawarrow[\E\ATBASE](\pmidx,\pmidy)
\put(\pmidx,5500){$b$}
\startphantom
\drawline\scalar[\S\REG](\particlefrontx,\particlefronty)[3]
\stopphantom
%
\drawline\fermion[\S\REG](\particlefrontx,\particlefronty)[\scalarlengthy]
\put(4500,\pmidy){$\Lambda^{a}_{G}$}
%
\drawline\fermion[\W\REG](\particlebackx,\particlebacky)[5500]
\drawarrow[\W\ATBASE](\pmidx,\pmidy)
\global\advance\pmidy by -2000
\put(\pmidx,\pmidy){$\bar{d}$}
%
\drawline\scalar[\E\REG](\particlefrontx,\particlefronty)[3]
\global\advance\pmidy by -2000
\global\advance\pmidx by -1000
\put(\pmidx,\pmidy){$Q^{J}_{(2)}$}
%
\drawline\fermion[\E\REG](\particlebackx,\particlebacky)[5500]
\drawarrow[\W\ATBASE](\pmidx,\pmidy)
\global\advance\pmidy by -2000
\put(\pmidx,\pmidy){$\bar{b}$}
\startphantom
\drawline\scalar[\S\REG](\particlefrontx,\particlefronty)[3]
\stopphantom
%
\drawline\fermion[\N\REG](\particlefrontx,\particlefronty)[\scalarlengthy]
\global\advance\pmidx by 1000
\put(\pmidx,\pmidy){$\tilde{z}^{\prime}$}
%
\drawline\fermion[\E\REG](\fermionbackx,\fermionbacky)[5500]
\drawarrow[\E\ATBASE](\pmidx,\pmidy)
\put(\pmidx,5500){$d$}
%
\thicklines
\drawline\fermion[\E\REG](30000,5000)[5500]
\global\advance\pmidx by -1000
\put(\pmidx,6000){$\tilde{z}^{\prime}$}
%
\drawline\fermion[\W\REG](\particlefrontx,\particlefronty)[5500]
\drawarrow[\E\ATBASE](\pmidx,\pmidy)
\put(\pmidx,5500){$b$}
%
\drawline\scalar[\S\REG](\particlefrontx,\particlefronty)[3]
\global\advance\pmidx by -4000
\put(\pmidx,\pmidy){$Q^{I}_{(2)}$}
%
\drawline\fermion[\W\REG](\particlebackx,\particlebacky)[5500]
\drawarrow[\W\ATBASE](\pmidx,\pmidy)
\global\advance\pmidy by -2000
\put(\pmidx,\pmidy){$\bar{d}$}
%
\drawline\fermion[\E\REG](\particlefrontx,\particlefronty)[5500]
\global\advance\pmidy by -2000
\global\advance\pmidx by -1000
\put(\pmidx,\pmidy){$\Lambda^{a}_{G}$}
%
\drawline\fermion[\E\REG](\particlebackx,\particlebacky)[5500]
\drawarrow[\E\ATBASE](\pmidx,\pmidy)
\global\advance\pmidy by -2000
\put(\pmidx,\pmidy){$d$}
%
\drawline\scalar[\N\REG](\particlefrontx,\particlefronty)[3]
\global\advance\pmidx by 1000
\put(\pmidx,\pmidy){$Q^{J}_{(2)}$}
%
\drawline\fermion[\E\REG](\scalarbackx,\scalarbacky)[5500]
\drawarrow[\W\ATBASE](\pmidx,\pmidy)
\put(\pmidx,5500){$\bar{b}$}
\end{picture}
\vskip 0.5in
\begin{picture}(10000,10000)(0, -2000)
%
\thicklines
\drawline\fermion[\E\REG](7000,5000)[5500]
\global\advance\pmidx by -1000
\put(\pmidx,6000){$\tilde{z}^{\prime}$}
%
\drawline\fermion[\W\REG](\particlefrontx,\particlefronty)[5500]
\drawarrow[\E\ATBASE](\pmidx,\pmidy)
\put(\pmidx,5500){$b$}
%
\drawline\scalar[\S\REG](\particlefrontx,\particlefronty)[3]
\put(3000,\pmidy){$Q^{I}_{(2)}$}
%
\drawline\fermion[\W\REG](\particlebackx,\particlebacky)[5500]
\drawarrow[\W\ATBASE](\pmidx,\pmidy)
\global\advance\pmidy by -2000
\put(\pmidx,\pmidy){$\bar{d}$}
%
\drawline\fermion[\E\REG](\particlefrontx,\particlefronty)[5500]
\global\advance\pmidy by -2000
\global\advance\pmidx by -1000
\put(\pmidx,\pmidy){$\Lambda^{a}_{G}$}
%
\drawline\fermion[\E\REG](\particlebackx,\particlebacky)[5500]
\drawarrow[\W\ATBASE](\pmidx,\pmidy)
\global\advance\pmidy by -2000
\put(\pmidx,\pmidy){$\bar{b}$}
%
\drawline\scalar[\N\REG](\particlefrontx,\particlefronty)[3]
\global\advance\pmidx by 1000
\put(\pmidx,\pmidy){$Q^{J}_{(2)}$}
%
\drawline\fermion[\E\REG](\scalarbackx,\scalarbacky)[5500]
\drawarrow[\E\ATBASE](\pmidx,\pmidy)
\put(\pmidx,5500){$d$}
%
\thicklines
\drawline\scalar[\E\REG](30000,5000)[3]
\global\advance\pmidx by -1000
\put(\pmidx,6000){$Q^{I}_{(2)}$}
%
\drawline\fermion[\W\REG](\particlefrontx,\particlefronty)[5500]
\drawarrow[\E\ATBASE](\pmidx,\pmidy)
\put(\pmidx,5500){$b$}
\startphantom
\drawline\scalar[\S\REG](\particlefrontx,\particlefronty)[3]
\stopphantom
%
\drawline\fermion[\S\REG](\particlefrontx,\particlefronty)[\scalarlengthy]
\global\advance\pmidx by -2500
\put(\pmidx,\pmidy){$\tilde{z}^{\prime}$}
%
\drawline\fermion[\W\REG](\particlebackx,\particlebacky)[5500]
\drawarrow[\W\ATBASE](\pmidx,\pmidy)
\global\advance\pmidy by -2000
\put(\pmidx,\pmidy){$\bar{d}$}
%
\drawline\scalar[\E\REG](\particlefrontx,\particlefronty)[3]
\global\advance\pmidy by -2000
\global\advance\pmidx by -1000
\put(\pmidx,\pmidy){$Q^{J}_{(2)}$}
%
\drawline\fermion[\E\REG](\particlebackx,\particlebacky)[5500]
\drawarrow[\W\ATBASE](\pmidx,\pmidy)
\global\advance\pmidy by -2000
\put(\pmidx,\pmidy){$\bar{b}$}
\startphantom
\drawline\scalar[\S\REG](\particlefrontx,\particlefronty)[3]
\stopphantom
%
\drawline\fermion[\N\REG](\particlefrontx,\particlefronty)[\scalarlengthy]
\global\advance\pmidx by 1000
\put(\pmidx,\pmidy){$\Lambda^{a}_{G}$}
%
\drawline\fermion[\E\REG](\fermionbackx,\fermionbacky)[5500]
\drawarrow[\E\ATBASE](\pmidx,\pmidy)
\put(\pmidx,5500){$d$}
\end{picture}
\vskip 0.5in
\begin{picture}(10000,10000)(0, -2000)
%
\thicklines
\drawline\fermion[\E\REG](7000,5000)[6000]
\global\advance\pmidx by -1000
\put(\pmidx,6000){$\Lambda^{a}_{G}$}
%
\drawline\fermion[\W\REG](\particlefrontx,\particlefronty)[5500]
\drawarrow[\E\ATBASE](\pmidx,\pmidy)
\put(\pmidx,5500){$b$}
%
\drawline\scalar[\SE\REG](\particlefrontx,\particlefronty)[4]
\global\advance\pmidx by 2000
\global\advance\pmidy by -1000
\put(\pmidx,\pmidy){$Q^{I}_{(2)}$}
%
\drawline\fermion[\E\REG](\particlebackx,\particlebacky)[5500]
\drawarrow[\E\ATBASE](\pmidx,\pmidy)
\global\advance\pmidy by -2000
\put(\pmidx,\pmidy){$d$}
%
\drawline\fermion[\W\REG](\particlefrontx,\particlefronty)[6000]
\global\advance\pmidy by -2000
\global\advance\pmidx by -1000
\put(\pmidx,\pmidy){$\tilde{z}^{\prime}$}
%
\drawline\fermion[\W\REG](\particlebackx,\particlebacky)[5500]
\drawarrow[\W\ATBASE](\pmidx,\pmidy)
\global\advance\pmidy by -2000
\put(\pmidx,\pmidy){$\bar{d}$}
%
\drawline\scalar[\NE\REG](\particlefrontx,\particlefronty)[4]
\global\advance\pmidx by -4000
\global\advance\pmidy by -1000
\put(\pmidx,\pmidy){$Q^{J}_{(2)}$}
%
\drawline\fermion[\E\REG](\scalarbackx,\scalarbacky)[5500]
\drawarrow[\W\ATBASE](\pmidx,\pmidy)
\put(\pmidx,5500){$\bar{b}$}
%
\thicklines
\drawline\scalar[\E\REG](30000,5000)[3]
\global\advance\pmidx by -1000
\put(\pmidx,6000){$Q^{I}_{(2)}$}
%
\drawline\fermion[\W\REG](\particlefrontx,\particlefronty)[5500]
\drawarrow[\E\ATBASE](\pmidx,\pmidy)
\put(\pmidx,5500){$b$}
%
\drawline\fermion[\SE\REG](\particlefrontx,\particlefronty)[7500]
\global\advance\pmidx by 2000
\global\advance\pmidy by -1000
\put(\pmidx,\pmidy){$\tilde{z}^{\prime}$}
%
\drawline\fermion[\E\REG](\particlebackx,\particlebacky)[5500]
\drawarrow[\W\ATBASE](\pmidx,\pmidy)
\global\advance\pmidy by -2000
\put(\pmidx,\pmidy){$\bar{b}$}
%
\drawline\scalar[\W\REG](\particlefrontx,\particlefronty)[3]
\global\advance\pmidy by -2000
\global\advance\pmidx by -1000
\put(\pmidx,\pmidy){$Q^{J}_{(2)}$}
%
\drawline\fermion[\W\REG](\particlebackx,\particlebacky)[5000]
\drawarrow[\W\ATBASE](\pmidx,\pmidy)
\global\advance\pmidy by -2000
\put(\pmidx,\pmidy){$\bar{d}$}
%
\global\advance\particlefrontx by 500
\drawline\fermion[\NE\REG](\particlefrontx,\particlefronty)[7500]
\global\advance\pmidx by -4000
\global\advance\pmidy by -1000
\put(\pmidx,\pmidy){$\Lambda^{a}_{G}$}
%
\drawline\fermion[\E\REG](\fermionbackx,\fermionbacky)[5500]
\drawarrow[\E\ATBASE](\pmidx,\pmidy)
\put(\pmidx,5500){$d$}
\end{picture}
\end{flushleft}
\label{fig1*p*3-5}
\end{figure}
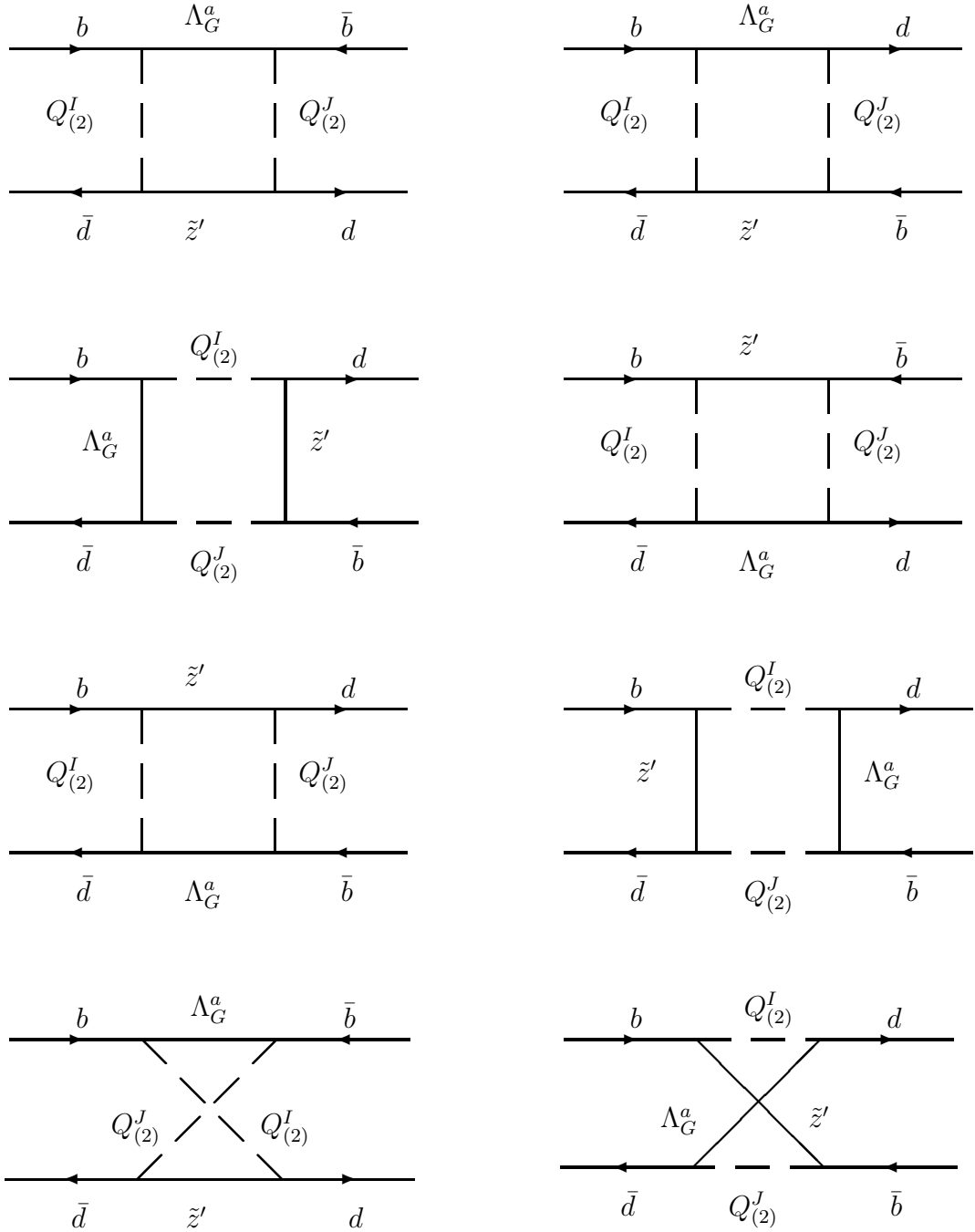
Since the standard model, $Sp(6)$ model and the MSSM are all part of SUSY
$Sp(6)$, we include their contributions in figures
\ref{fig1*p*3-4}, \ref{fig2*p*3-4} and \ref{fig3*p*3-4}. Let us turn our
attention to figures \ref{fig3*p*3-4} and \ref{fig1*p*3-5}.

To get the physical fields for the down squarks, we use the following
approximate equations,
\begin{equation}
\label{eq1a*p*3-5}
Q^{\prime I}_{2 \alpha} = {\displaystyle \sum_{J}} V^{I J}_{u} Q^{J}_{2 \alpha}
\end{equation}
\begin{equation}
Q^{\prime I}_{1 \alpha} = {\displaystyle \sum_{J}} V^{I J}_{u} Q^{J}_{1 \alpha}
\end{equation}
\begin{equation}
D^{\prime I}_{\alpha} = {\displaystyle \sum_{J}} V^{I J}_{D} D^{J}_{\alpha}
\end{equation}
\begin{equation}
\label{eq1d*p*3-5}
U^{\prime I}_{\alpha} = {\displaystyle \sum_{J}} V^{I J}_{U} U^{J}_{\alpha}
\end{equation}
where $V_{u}$, $V_{D}$ and $V_{U}$ are the unitary matrices in the MSSM which
redefine the initial fermion and sfermion fields to get
the corresponding physical fields as in
\begin{equation}
\label{eq1a*p*3-6}
\left( Q^{\prime I}_{(i_{2}) \alpha},\;\psi^{\prime I}_{q\, (i_{2}) \alpha}
\right) = {\displaystyle \sum_{J}} V^{I J}_{Q (i_{2})} \left(
       Q^{J}_{(i_{2}) \alpha},\;\psi^{J}_{q\, (i_{2}) \alpha} \right)
\end{equation}
\begin{equation}
\left( U^{\prime I}_{\alpha},\;\psi^{\prime I}_{U \alpha} \right) =
{\displaystyle \sum_{J}} V^{I J}_{U} \left(
       U^{J}_{\alpha},\;\psi^{J}_{U \alpha} \right)
\end{equation}
\clearpage
\begin{equation}
\label{eq1c*p*3-6}
\left( D^{\prime I}_{\alpha},\;\psi^{\prime I}_{D \alpha} \right) =
{\displaystyle \sum_{J}} V^{I J}_{D} \left(
       D^{J}_{\alpha},\;\psi^{J}_{D \alpha} \right).
\end{equation}
Here $i_{2} = 1,\,2$ for the \underline{2} representation of $SU_{L}(2)$ and
$V_{Q 1} = V_{u}$, $V_{Q 2} = V_{d}$.

Note that equation \ref{eq1a*p*3-5} is peculiar when compared to equation
\ref{eq1a*p*3-6} (with $i_{2} = 2$  and $V_{Q (2)} = V_{d}$)
because instead of having $V_{d}$, we have $V_{u}$. Equations \ref{eq1a*p*3-5}
to \ref{eq1d*p*3-5} were derived under the following assumptions:
\begin{enumerate}
  \item we neglect $Q^{\prime}_{1}\!-\!U^{\prime}$ and
$Q^{\prime}_{2}\!-\!D^{\prime}$ mixing;
  \item we let $(m^{2}_{D})^{IJ} \longrightarrow (m^{2}_{D})^{I} \delta^{IJ}$,
        $(m^{2}_{U})^{IJ} \longrightarrow (m^{2}_{U})^{I} \delta^{IJ}$ (no sum
in $I$) and
  \item the evolution of the renormalization group equations down to low
energies results in
        \begin{equation}
         M^{2}_{Q_{2}} \simeq \mu^{(0)}_{Q_{2}} {\bf 1} + \mu^{(2)}_{Q_{2}}
M^{\dagger}_{q_{u}} M_{q_{u}}
        \end{equation}
        \begin{equation}
         M^{2}_{Q_{1}} \simeq \mu^{(0)}_{Q_{1}} {\bf 1} + \mu^{(1)}_{Q_{1}}
M^{\dagger}_{q_{u}} M_{q_{u}}
        \end{equation}
        \begin{equation}
         M^{2}_{D} \simeq \mu^{(0)}_{D} {\bf 1}
        \end{equation}
        \begin{equation}
         M^{2}_{U} \simeq \mu^{(0)}_{U} {\bf 1} + \mu^{(1)}_{U} M_{q_{u}}
M^{\dagger}_{q_{u}}.
        \end{equation}
\end{enumerate}

Item 1 above is a usual assumption made \cite{Gunion}, \cite{Franzini} and
\cite{Gerard}. In some low-energy supergravity models, this mixing tends to be
small. However, sometimes this assumption maybe inadequate \cite{Gunion}.

Item 2 comes from the usual assumption made that all SUSY scalars have
degenerate mass \cite{Gabbiani}.

In item 3, we \underline{neglected} terms proportional to $M^{\dagger}_{q_{d}}
M_{q_{d}}$. This is justifiable for some supergravity models where a large
top quark mass is responsible for $SU_{L}(2) \times U_{Y}(1)$ breaking in the
low energy effective theory. \cite{Gunion}.

We are also inherently making the approximation in the $Sp(6)$ model that the
initial fermion fields are transformed into physical fields by the same
transformation matrices $V_{u}$, $V_{d}$, $V_{U}$ and $V_{D}$ which were used
in rotating the initial fields to physical fields in the SM as in
equations \ref{eq1a*p*3-6} to \ref{eq1c*p*3-6}. We justify this by considering
the $Sp(6)$ model when it has already spontaneously broken down
to the SM.

Similar to the discussion in reference \cite{Blad-Kuo}, the mixing parameter
$x_{d}$ is
\begin{equation}
\label{eq1*p*3-7}
x_{d} \simeq \frac{2 \left| M_{1 2} \right|}{\Gamma}
\end{equation}
where
\begin{equation}
\label{eq2*p*3-7}
M_{12}=\left< \bar{B}^{0} |{\cal H}_{eff}| B^{0} \right>
\end{equation}
and $\Gamma = \frac{1}{\tau_{B}}$ where $\tau_{B}$ is the mean lifetime of the
$B^{0}$ meson.

Adding the contributions due to all the diagrams in figures \ref{fig1*p*3-4},
\ref{fig2*p*3-4}, \ref{fig3*p*3-4} and \ref{fig1*p*3-5}, we get the following
$\cal H \mit _{eff}$,
\begin{equation}
\cal H \mit _{eff} = F \rm
\bar{\Psi}^{d}_{\alpha}\gamma_{\mu}\frac{1}{2}(1-\gamma^{5})\Psi^{b}_{\alpha}
\bar{\Psi}^{d}_{\beta}\gamma^{\mu}\frac{1}{2}(1-\gamma^{5})\Psi^{b}_{\beta}
\label{eq3*p*3-7}
\end{equation}
where the operators are understood to be normal ordered and $F$ is defined as
\begin{equation}
\label{eq1*p*3-8}
\begin{array}{lll}
F & \equiv & \frac{G_{F}^{2}}{4\pi^{2}}m_{t}^{2}\frac{A(z_{t})}{z_{t}} \left[
\left( V^{\dagger} \right)^{1 3} \left( V \right)^{3 3}\right]^{2}
\eta_{QCD} + \frac{9 \sqrt{2}}{4} G_{F} \left( \frac{m_{W}}{m_{Z^{\prime}}}
\right)^{2}
\left[  \left( V^{\dagger}_{d} \right)^{1 3} \left( V_{d} \right)^{3 3}
\right]^{2} \eta^{sp} \\
 &  & \frac{\alpha^{2}_{s}}{36 m^{2}_{\tilde{g}}} {\displaystyle \sum_{I J}}
\left( V^{\dagger}  \right)^{1 I} V^{I 3}
                              \left( V^{\dagger}  \right)^{1 J} V^{J 3} \left[
4 \cal I \mit _{I J} + 11 \cal K \mit _{I J} \right] \\
 &  & \frac{\sqrt{2} G_{F} \alpha_{s}}{24 \pi} \left(
\frac{m_{W}}{m_{\tilde{g}}}  \right)^{2} {\displaystyle \sum_{I J}}
                                                \left(

\frac{m_{\tilde{z}^{\prime}}}{m_{\tilde{g}}}
                                                     \left[V^{I 3} V^{J 3} \cal
V \mit ^{I}_{sp} \cal V \mit ^{J}_{sp}
                                               + \left( V^{\dagger}
\right)^{\rm 1 \mit I} \left( V^{\dagger} \right)^{\rm 1 \mit J} \cal T
                                                  \mit ^{I}_{sp} \cal T \mit
^{J}_{sp}
                                                     \right] \cal I \mit_{M I
J}
                                                \right. \\
  &  & \left. + \, V^{I 3} \left( V^{\dagger} \right)^{1 J} \cal V \mit
^{I}_{sp} \cal T \mit ^{J}_{sp} \cal K \mit_{M I J} \right)
\end{array}
\end{equation}
where $V$ is the CKM matrix as in reference \cite{Franzini}. We will use
\begin{equation}
V=\left( \begin{array}{ccc}
V_{ud} & V_{us} & V_{ub} \\
V_{cd} & V_{cs} & V_{cb} \\
V_{td} & V_{ts} & V_{tb}
\end{array} \right )\simeq
\left( \begin{array}{ccc}
1-\frac{1}{2}\lambda^{2}        & \lambda                  &
A\lambda^{3}\rho e^{i\phi} \\
-\lambda                        & 1-\frac{1}{2}\lambda^{2} &
A\lambda^{2} \\
A\lambda^{3}(1-\rho e^{-i\phi}) & -A\lambda^{2}            &
1
\end{array} \right).
\end{equation}

In equation \ref{eq1*p*3-8}, $G_{F}$ is the Fermi constant, $m_{t}$ is the top
mass, $\eta_{QCD} \simeq 0.85$ is the QCD correction for the SM
graphs in figure \ref{fig1*p*3-4}, $m_{Z^{\prime}}$ is the $Z^{\prime}$ mass,
$\alpha_{s} = \frac{g_{3}^{2}}{4 \pi} \simeq 0.1134$,
$m_{\tilde{g}}$, is the gluino mass, $m_{\tilde{z}^{\prime}}$ is the
$Z^{\prime}$-ino mass and $m_{W}$ is the W-boson mass. $\eta^{sp}$
is the QCD correction to the tree level graph in figure \ref{fig2*p*3-4} and is
given by \cite{Blad-Kuo},
\begin{equation}
\eta^{sp}=
\left[\frac{\alpha_{s}(m_{Z^{\prime}}^{2})}{\alpha_{s}(m_{t}^{2})}\right]^{6/21}
\left[\frac{\alpha_{s}(m_{t}^{2})}
{\alpha_{s}^{\prime}(\mu^{2})}\right]^{6/23}
\end{equation}
where the running strong coupling constant is
\begin{equation}
\alpha_{s}(Q^{2})=\frac{12\pi}{33-2n_{f}}
\frac{1}{ln(\frac{Q^{2}}{\Lambda^{2}})}.
\end{equation}

Note that $n_{f}$ is the number of quark flavors and $\alpha_{s}^{\prime}$ is
$\alpha_{s}$ evaluated in an effective five-quark theory
resulting from the step of removing the t-quark from explicitly appearing in
the theory.

$\frac{A(z_{t})}{z_{t}}$ is given by \cite{Blad-Kuo},
\begin{equation}
\frac{A(z_{t})}{z_{t}}=\frac{1}{4}+\frac{9}{4(1-z_{t})}-
\frac{3}{2(1-z_{t})^{2}}-\frac{3z_{t}^{2}\ln z_{t}}{2(1-z_{t})^{3}}\;
\end{equation}
where
\begin{equation}
z_{t}\equiv (m_{t}/m_{W})^{2}.
\end{equation}

The functions $\cal I \mit _{I J}$, $\cal K \mit _{I J}$, $\cal V \mit
^{I}_{sp}$, $\cal T \mit ^{I}_{sp}$, $\cal I \mit _{M I J}$ and
$\cal K \mit _{M I J}$ are given by
\begin{equation}
\cal I \mit _{I J} \equiv \frac{1}{z_{I}-z_{J}} \left \{ \left[ \frac{z_{I} \ln
z_{I}}{(1 - z_{I})^{2}} + \frac{1}{1 - z_{I}} \right] -
                           \left[ \frac{z_{J} \ln z_{J}}{(1 - z_{J})^{2}} +
\frac{1}{1 - z_{J}} \right] \right \}
\end{equation}
\begin{equation}
\cal K \mit _{I J} \equiv \frac{1}{z_{I}-z_{J}} \left \{ \left[ \frac{z^{2}_{I}
\ln z_{I}}{(1 - z_{I})^{2}} + \frac{1}{1 - z_{I}} \right] -
                           \left[ \frac{z^{2}_{J} \ln z_{J}}{(1 - z_{J})^{2}} +
\frac{1}{1 - z_{J}} \right] \right \}
\end{equation}
\begin{equation}
\cal V \mit ^{I}_{sp} \equiv \left( V^{\dagger}  \right)^{\rm 1 \mit I} - \rm 3
\mit \left( V^{\dagger}_{d}  \right)^{\rm 1 3}  V_{u}^{\rm 3 \mit I}
\end{equation}
\begin{equation}
\cal T \mit ^{I}_{sp} \equiv  V^{I \rm 3} - \rm 3 \mit \left( V^{\dagger}_{u}
\right)^{I \rm 3} V^{\rm 3 3}_{d}
\end{equation}
\begin{equation}
\begin{array}{ccl}
\cal I \mit_{M I J} & \equiv & \frac{z_{M} \ln z_{M}}{(z_{M} - 1) (z_{M} -
z_{I}) (z_{M} - z_{J})} + \frac{z_{I} \ln z_{I}}{(z_{I} - 1) (z_{I} - z_{J})
(z_{I} - z_{M})} \\
                    &        & +\, \frac{z_{J} \ln z_{J}}{(z_{J} - 1) (z_{J} -
z_{I}) (z_{J} - z_{M})}
\end{array}
\end{equation}
\begin{equation}
\begin{array}{ccl}
\cal K \mit_{M I J} & \equiv & \frac{z_{M}^{2} \ln z_{M}}{(z_{M} - 1) (z_{M} -
z_{I}) (z_{M} - z_{J})} + \frac{z_{I}^{2} \ln z_{I}}{(z_{M} - 1) (z_{I} -
z_{J}) (z_{I} - z_{M})}\\
                    &        & \frac{z_{J}^{2} \ln z_{J}}{(z_{J} - 1) (z_{J} -
z_{I}) (z_{J} - z_{M})}
\end{array}
\end{equation}
where
\begin{equation}
z_{I} \equiv \left( \frac{m^{I}_{Q_{2}}}{m_{\tilde{g}}}  \right)^{2}
\end{equation}
\begin{equation}
z_{M} \equiv \left( \frac{m_{\tilde{z}^{\prime}}}{m_{\tilde{g}}}  \right)^{2}
\end{equation}
$m^{I}_{Q_{2}}$ here are the down squark masses given by
\begin{equation}
m^{1}_{Q_{2}} = m_{\tilde{d}} = \sqrt{m^{2}_{\tilde{b}} + \left|
\mu^{(2)}_{Q_{2}}  \right| m^{2}_{t}}
\end{equation}
\begin{equation}
m^{2}_{Q_{2}} = m_{\tilde{s}} = m_{\tilde{d}}
\end{equation}
\begin{equation}
m^{3}_{Q_{2}} = m_{\tilde{b}}\;.
\end{equation}

Note that in deriving the $\cal H \mit _{eff}$ of equation \ref{eq3*p*3-7}, the
gluinos and the $\tilde{z}^{\prime}$ are majorana spinors. We
assign for the four-majorana spinors,
\begin{equation}
\Lambda^{a}_{G} = \left( \begin{array}{c}
                   - i \lambda^{a}_{G} \\
                   i \bar{\lambda}^{a}_{G}
                       \end{array} \right)
\end{equation}
\begin{equation}
\tilde{z}^{\prime} = \left( \begin{array}{c}
                   - i \lambda_{Z^{\prime}} \\
                   i \bar{\lambda}_{Z^{\prime}}
                       \end{array} \right)
\label{eq1b*p*3-10suppl1}
\end{equation}
Since the gluinos and the $Z^{\prime}$-inos are majorana spinors, the Feynman
rules needed to deal with them are tricky. There had been a
number of papers ~\cite{Haber-Kane} and \cite{Jones}, which discuss Feynman
rules for Majorana spinors. The best and the most recent paper which
we used in deriving $\cal H \mit _{eff}$ in equation~\ref{eq3*p*3-7}  is
reference~\cite{Denner}.

In equation \ref{eq1b*p*3-10suppl1}, we assume $\tilde{z}^{\prime}$ to be the
physical field. We do this to make the calculations more manageable
and to do away with too many arbitrary parameters. Strictly speaking, the
formation of this neutralino involves the mixing of the additional neutral
higgsinos and gauginos.

We note here that similar to equation \ref{eq5*p*2-4}, we have for the
$\tilde{z}^{\prime}$ in equation \ref{eq1b*p*3-10suppl1}
\begin{equation}
\lambda_{Z^{\prime}} = \lambda_{A(\rm 18)}
\end{equation}
i.e. it is a component of the superfield $\cal A \mit _{\rm 18}$ in the
lagrangian of section \ref{susy-sp1}.

Using equation \ref{eq3*p*3-7}  in equation \ref{eq2*p*3-7} we can calculate
$x_{d}$ for SUSY $Sp(6)$ using equation \ref{eq1*p*3-7}.
We get
\begin{equation}
\label{eq2*p*3-10}
x_{d} = \frac{2}{3} \left[ B_{B} f_{B}^{2}  \right] M_{B} \tau_{B_{d}} \left| F
\right|
\end{equation}
where $F$ is given by equation \ref{eq1*p*3-8}. In equation \ref{eq2*p*3-10},
we applied the vacuum insertion approximation using the normalization for the
$\pi \longrightarrow \mu \nu$
\begin{equation}
\left< 0 |\bar{\psi}_{u}\gamma_{\mu}\gamma^{5}\psi_{d}|\pi \right>=
\frac{ip_{\mu}f_{\pi}}{\sqrt{2E_{p}}}.
\end{equation}

We  have  the ``bag'' factor $B_{B}$ which takes into account all deviations
from the vacuum insertion approximation. The quantity $f_{B}$ is the
corresponding $f_{\pi}$ for the $B$ meson system. $M_{B}$ is the mass of the
$B^{0}$ meson. We note here that coefficients due to color statistics as in
reference \cite{Hagelin} for the various graphs have been
carefully calculated following the discussion of section II of reference
\cite{kelley}.

With equation \ref{eq2*p*3-10}, we can demonstrate that the
$\tilde{z}^{\prime}$ contribution can play a significant role in suppressing
the mixing parameter $x_{d}$ to within the experimentally acceptable range
$0.57 \stackrel{<}{\sim} x_{d} \stackrel{<}{\sim} 0.77$
\cite{Schroder} especially now that the top quark mass can assume a large value
of $158 \stackrel{<}{\sim} m_{t} \stackrel{<}{\sim} 194$ Gev.
 Using reasonable values of the parameters, $m_{\tilde{g}} = 141 $ Gev,
$m_{Z^{\prime}} = 4$ Tev, $m_{\tilde{z}^{\prime}} = 900 $ Gev,
we present the plots in figure \ref{fg1*p*3-11}. For the uncertain parameters
$B_{B} f_{B}^{2}$ and the CKM matrix's $A$ and $\rho$, we used the central
values. $\phi$ in the CKM matrix was set to $\frac{\pi}{2}$.
\begin{figure}[htbp]
\centering
  \caption{Plot of the $B^{0}_{d}\bar{B}^{0}_{d}$ mixing parameter $x_{d}$
versus the top mass $m_{t}$ for the SM (solid line), MSSM (dashed-line) and for
SUSY
$Sp(6)$ (dot-dashed line). The region between the two \underline{horizontal
lines} are the experimentally
allowed region from the ARGUS, CLEO result.}
  \label{fg1*p*3-11}
\end{figure}

The region between the two horizontal lines are the experimentally allowable
range for the mixing parameter $x_{d}$.
Figure \ref{fg1*p*3-11}
seem to indicate that the standard model and the minimal supersymmetric
standard model may not be appealing since $x_{d}$ is a bit too high
for the experimentally allowable range for $158 \stackrel{<}{\sim} m_{t}
\stackrel{<}{\sim} 194$ Gev.
\footnote{Of course, these models can still give low values of $x_{d}$ for some
range of the uncertain parameters.
However, as indicated above, we used the central values of these uncertain
parameters which we feel is the more
reasonable thing to do.} With the inclusion of the $Z^{\prime}$
and $\tilde{z}^{\prime}$ contributions, however, a \underline{suppression}
(mainly due to the $\tilde{z}^{\prime}$)  of \underline{$x_{d}$}
occurs which makes it fall well within the allowable range for $158
\stackrel{<}{\sim} m_{t} \stackrel{<}{\sim} 194$ Gev. The $\tilde{z}^{\prime}$
contribution inherently cancels
out the other contributions properly to make $x_{d}$ fall within the
experimentally preferred range. Note that we used $m_{Z^{\prime}} = 4$ Tev
here.
It turns out that because of the suppression due to the $\tilde{z}^{\prime}$, a
$m_{Z^{\prime}} = 3$ Tev can still yield an $x_{d}$ within the allowable region
for
$158 \stackrel{<}{\sim} m_{t} \stackrel{<}{\sim} 194$ Gev. Although in
figure~\ref{fg1*p*3-11} we did
not include QCD corrections on the SUSY graphs, we expect them to suppress
$x_{d}$ further, allowing for
even lighter $m_{Z^{\prime}}$ of about $2$ Tev.

A very interesting feature of the $\tilde{z}^{\prime}$ contribution is that
$m_{\tilde{z}^{\prime}} < m_{Z^{\prime}}$ if the cancellation is to be big
enough.
We have here a model where experiment indicates the possibility of a gaugino
with less  mass than the gauge boson.

On page 637 of reference~\cite{Bertolini}, it was mentioned that neutralinos
$\kappa^{0}_{i}$ in MSSM (actually a subclass of this) contribute negligibly
to FCNC amplitudes. Here, however, we have a neutralino, namely
$\tilde{z}^{\prime}$, which may contribute significantly to FCNC in the form of
$x_{d}$.
The reason is again unique to the presence of the horizontal symmetry.
Couplings due to $\kappa^{0}_{i}$ result to factors $\sim\; G^{2}_{F}$
while for $\tilde{z}^{\prime}$ one gets factors $\sim\; G_{F} \alpha_{s}$.
These arise only if there is a (weak) neutralino and a gluino
traversing the loop of the box graph. Since the gluino cannot change flavor,
the neutralino must. Hence, only a theory with flavor-changing
(weak) neutralinos can have this such as in the SUSY $Sp(6)$ model.

A number of recent papers~\cite{Dine-L} to \cite{Nir}, discussed flavor (or
horizontal) symmetries in SUSY in the context of squark mass degeneracy.
One motivation for this is that squark mass degeneracy tends to control large
SUSY FCNC effects. In what we have just presented, we can view flavor
symmetries in SUSY in another light, and that is, the presence of the SUSY
partners (like $\tilde{z}^{\prime}$) of the extra horizontal gauge
bosons (like $Z^{\prime}$) of a supersymmetric non-abelian horizontal gauge
theory may further suppress FCNC in SUSY theories
by cancelling the other contributions as demonstrated by the
SUSY $Sp(6)$ model.
\clearpage
\section{Conclusions and Outlook}
\label{conclude}

The $Sp(6)$ model has been a very interesting model in addressing the
generation problem. The presence of the horizontal subgroup $SU_{H}(3)$
which relates the different generations gives rise to extra gauge bosons, the
lightest set of which are the
$\left(  W_{1}^{\prime},\,W_{2}^{\prime}, Z^{\prime} \right)$. So far
phenomenological studies have been concentrated on the $Z^{\prime}$
effects to FCNC.

Because of the recent renewed interest on SUSY theories, we have presented in
this paper a supersymmetric extension of the $Sp(6)$
model (SUSY $Sp(6)$) by writing down a supersymmetric $SU_{C}(3) \times
Sp_{L}(6) \times U_{Y}(1)$ gauge invariant lagrangian. Its derivation
follows closely that of the extension of the
standard model to the minimal supersymmetric standard model. With the
introduction of the second type of higgs,
we can easily make SUSY  $Sp(6)$ anomaly-free as in the MSSM.

As a first step to studying the phenomenological consequences of SUSY  $Sp(6)$,
we analyzed its dominant contributions to
$B^{0}_{d} \bar{B}^{0}_{d}$ mixing which include that of the SM, $Sp(6)$ model,
MSSM and the contributions due to the $Z^{\prime}$-ino
($\tilde{z}^{\prime}$). By plotting the mixing parameter $x_{d}$ (as computed
in the SUSY  $Sp(6)$ framework) versus $m_{t}$ (top mass),
we are led to the following observations:
\begin{enumerate}
   \item a cancellation of FCNC effects due to $\tilde{z}^{\prime}$,
   \item more pronounced contribution of a weak neutralino in box graphs
involving FCNC and
   \item a gaugino with a lighter mass than the gauge boson.
\end{enumerate}

The first item can be a reason to possibly view studies of SUSY horizontal
symmetries in a new light. Instead of just looking at horizontal
symmetries in SUSY as a way to make squark mass degeneracy more natural thereby
reducing FCNC, we can also view horizontal symmetries in SUSY
with respect to the effects of the $Z^{\prime}$-ino which may reduce FCNC as
well.

The second item is fairly unique to SUSY theories with horizontal symmetries.
Weak neutralinos have usually negligible contributions in box graphs
since their coupling introduces $G_{F}^{2}$ whereas gluino graphs introduce
$\alpha_{s}^{2}$. With the weak neutralino $\tilde{z}^{\prime}$
which may change flavor, box graphs involving these can have bigger
contributions $\sim G_{F}\alpha_{s}$.

The third item can motivate further studies on SUSY $Sp(6)$ since the
$m_{\tilde{z}^{\prime}}<m_{Z^{\prime}}$ and thus the effects of
$\tilde{z}^{\prime}$ may be more accessible with respect to the accelerator
energies at present.

With the formulation of SUSY  $Sp(6)$, one can use the workable lagrangian
which we have written down (and which we have checked to reproduce
known results in SM, MSSM and $Sp(6)$), to study other phenomenological
consequences of a supersymmetric $Sp(6)$ model.

\clearpage
\begin{center}
{\Large \bf Appendix}
\end{center}
\clearpage
\section{Conventions, Notation and Formulas}

    \begin{equation}
g_{\mu \nu} =  \left( \begin{array}{cccc}
1 & 0   & 0  & 0 \\
0 & -1  & 0  & 0 \\
0 & 0   & -1 & 0 \\
0 & 0   & 0  & -1
\end{array} \right)
     \end{equation}

    \begin{equation}
\eta_{i_{6}\; j_{6}} = \left( \begin{array}{cccrrr}
0 & 0 & 0 & -1 & 0 & 0   \\
0 & 0 & 0 & 0 & -1 & 0   \\
0 & 0 & 0 & 0 & 0 & -1   \\
1 & 0 & 0 & 0 & 0 & 0   \\
0 & 1 & 0 & 0 & 0 & 0   \\
0 & 0 & 1 & 0 & 0 & 0
\end{array} \right)
     \end{equation}

    \begin{equation}
         Q = T_{3} + \frac{y}{2}
     \end{equation}

    \begin{equation}
\begin{array}{ccccc}

T^{(1)} = \frac{1}{2 \sqrt{2}} \sigma_{1} \otimes \lambda^{0} & ;
&
T^{(2)} = \frac{1}{2 \sqrt{2}} \sigma_{1} \otimes \lambda^{1} & ;
&
T^{(3)} = \frac{1}{2 \sqrt{2}} \sigma_{1} \otimes \lambda^{3}
\\

T^{(4)} = \frac{1}{2 \sqrt{2}} \sigma_{1} \otimes \lambda^{4} & ;
&
T^{(5)} = \frac{1}{2 \sqrt{2}} \sigma_{1} \otimes \lambda^{6} & ;
&
T^{(6)} = \frac{1}{2 \sqrt{2}} \sigma_{1} \otimes \lambda^{8}
\\

T^{(7)} = \frac{1}{2 \sqrt{2}} \sigma_{2} \otimes \lambda^{0} & ;
&
T^{(8)} = \frac{1}{2 \sqrt{2}} \sigma_{2} \otimes \lambda^{1} & ;
&
T^{(9)} = \frac{1}{2 \sqrt{2}} \sigma_{2} \otimes \lambda^{3}
\\

T^{(10)} = \frac{1}{2 \sqrt{2}} \sigma_{2} \otimes \lambda^{4} & ;
&
T^{(11)} = \frac{1}{2 \sqrt{2}} \sigma_{2} \otimes \lambda^{6} & ;
&
T^{(12)} = \frac{1}{2 \sqrt{2}} \sigma_{2} \otimes \lambda^{8}
\\

T^{(13)} = \frac{1}{2 \sqrt{2}} \sigma_{3} \otimes \lambda^{0} & ;
&
T^{(14)} = \frac{1}{2 \sqrt{2}} \sigma_{3} \otimes \lambda^{1} & ;
&
T^{(15)} = \frac{1}{2 \sqrt{2}} \sigma_{3} \otimes \lambda^{3}
\\

T^{(16)} = \frac{1}{2 \sqrt{2}} \sigma_{3} \otimes \lambda^{4} & ;
&
T^{(17)} = \frac{1}{2 \sqrt{2}} \sigma_{3} \otimes \lambda^{6} & ;
&
T^{(18)} = \frac{1}{2 \sqrt{2}} \sigma_{3} \otimes \lambda^{8}
\\

T^{(19)} = \frac{1}{2 \sqrt{2}} {\bf 1} \otimes \lambda^{2} & ;
&
T^{(20)} = \frac{1}{2 \sqrt{2}} {\bf 1} \otimes \lambda^{5} & ;
&
T^{(21)} = \frac{1}{2 \sqrt{2}} {\bf 1} \otimes \lambda^{7}
\end{array}
\label{eq3*p*ApA-1}
     \end{equation}

\begin{equation}
[Y^{a},Y^{b}] = i f_{abc}\, Y^{c}, \rm where\;\; \mit Y^{a} =
\frac{\lambda^{a}}{2}= \; \mit SU_{C}(\rm 3)\;\; \rm
                                  generators\;\; in\;\; \underline{3}\;\;
representation
 \end{equation}

\begin{equation}
\bar{Y}^{a} = SU_{C}(\rm 3)\;\; \rm  generators\;\; in\;\;
\underline{\bar{3}}\;\; representation \ni  (\bar{\mit Y}^{a})_{\beta \alpha} =
          (-\mit Y^{a})_{\alpha \beta }
 \end{equation}

\begin{equation}
i, j, k = 1, 2, \ldots, 21 =\; Sp_{L}(6)\;\; \rm generator\;\; indices \;\;
unless \;\; specified
\;\; otherwise
\end{equation}

\begin{equation}
i_{6}, j_{6}, k_{6} = 1, 2, \ldots, 6 =\; \rm indices\;\; for\;\; the\;\;
\underline{6}\;\; \rm representation\;\;
                          of\;\; \mit Sp_{L}(\rm 6)
\end{equation}

\begin{equation}
I, J, K = 1, 2, 3 =\;  \rm generation\;\; indices
\end{equation}

\begin{equation}
\alpha, \beta, \gamma = 1, 2, 3 =\; \rm indices\;\; for\;\; the\;\;
\underline{3}\;\; \rm representation\;\;
                          of\;\; \mit SU_{C}(\rm 3)
\end{equation}

\begin{equation}
a, b, c = 1, 2, ..., 8 =\; SU_{C}(3)\;\; \rm generator\;\; indices
\end{equation}

\begin{equation}
g_{1},\;g_{2},\;g_{3}\; =\; U_{Y}(1),\;SU_{L}(2),\;SU_{C}(3)\;\; \rm gauge\;\;
couplings \;\; respectively
\end{equation}

\begin{equation}
B^{\nu},\;A^{\nu}_{i},\;G^{\mu}_{a}\; \equiv U_{Y}(
1),\;Sp_{L}(6),\;SU_{C}(3)\;\; \rm gauge\;\; bosons\;\; respectively
\end{equation}

\begin{equation}
\Psi_{L} \equiv \frac{1}{2} (1 - \gamma_{5}) \Psi
\;\;;\;\;
\Psi_{rt} \equiv \frac{1}{2} (1 + \gamma_{5}) \Psi
\end{equation}

\begin{equation}
\begin{array}{ccccc}
\Psi^{\prime}_{(1) L} & = & \nu^{\prime}_{e L} & = & \nu^{\prime 1}_{e L} \\
\Psi^{\prime}_{(2) L} & = & \nu^{\prime}_{\mu L} & = & \nu^{\prime 2}_{e L} \\
\Psi^{\prime}_{(3) L} & = & \nu^{\prime}_{\tau L} & = & \nu^{\prime 3}_{e L} \\
\Psi^{\prime}_{(4) L} & = & e^{\prime}_{L} & = & e^{\prime 1}_{L} \\
\Psi^{\prime}_{(5) L} & = & \mu^{\prime}_{L} & = & e^{\prime 2}_{L} \\
\Psi^{\prime}_{(6) L} & = & \tau^{\prime}_{L} & = & e^{\prime 3}_{L} \\
\end{array}
\end{equation}

\begin{equation}
\begin{array}{ccccc}
(\Psi^{\prime}_{Q})_{(1) \alpha L} & = & u^{\prime}_{\alpha L} & = & u^{\prime
1}_{\alpha L} \\
(\Psi^{\prime}_{Q})_{(2) \alpha L} & = & c^{\prime}_{\alpha L} & = & u^{\prime
2}_{\alpha L} \\
(\Psi^{\prime}_{Q})_{(3) \alpha L} & = & t^{\prime}_{\alpha L} & = & u^{\prime
3}_{\alpha L} \\
(\Psi^{\prime}_{Q})_{(4) \alpha L} & = & d^{\prime}_{\alpha L} & = & d^{\prime
1}_{\alpha L} \\
(\Psi^{\prime}_{Q})_{(5) \alpha L} & = & s^{\prime}_{\alpha L} & = & d^{\prime
2}_{\alpha L} \\
(\Psi^{\prime}_{Q})_{(6) \alpha L} & = & b^{\prime}_{\alpha L} & = & d^{\prime
3}_{\alpha L}
\end{array}
\end{equation}

\begin{equation}
\Psi^{1}_{rt} = e_{R}\;,\; \Psi^{2}_{rt} = \mu_{R}\;,\; \Psi^{3}_{rt} =
\tau_{R}
\end{equation}

\begin{equation}
(\Psi_{u})^{I}_{\alpha rt}\;,\;\ni \; (\Psi_{u})^{1}_{\alpha rt} = u_{\alpha R}
\;,\; (\Psi_{u})^{2}_{\alpha rt} = c_{\alpha R} \;,\;
                               (\Psi_{u})^{3}_{\alpha rt} = t_{\alpha R}
\end{equation}

\begin{equation}
(\Psi_{d})^{I}_{\alpha rt}\;,\;\ni \; (\Psi_{d})^{1}_{\alpha rt} = d_{\alpha R}
\;,\; (\Psi_{d})^{2}_{\alpha rt} = s_{\alpha R} \;,\;
                               (\Psi_{d})^{3}_{\alpha rt} = b_{\alpha R}
\end{equation}

\begin{equation}
\cal D \mit ^{(1) \mu} \Psi^{I}_{rt} \equiv \partial^{\mu} \Psi^{I}_{rt}
                                       + i \frac{g_{1}}{2} B^{\mu} y
\Psi^{I}_{rt}
\end{equation}

\begin{equation}
\cal D \mit ^{(2) \mu} \Psi^{\prime}_{(i_{6}) L} \equiv \partial^{\mu}
\Psi^{\prime}_{(i_{6}) L} + i g_{sp} A^{\mu}_{j} T^{(j)}_{i_{6} j_{6}}
\Psi^{\prime}_{(j_{6}) L}
                                       + i \frac{g_{1}}{2} B^{\mu} y
\Psi^{\prime}_{(i_{6}) L}
\end{equation}

\begin{equation}
\begin{array}{ccl}
\nabla^{\mu} (\Psi^{\prime}_{Q})_{(i_{6}) \alpha L} & \equiv & \partial^{\mu}
(\Psi^{\prime}_{Q})_{(i_{6}) \alpha L}
+ i g_{3} G^{\mu}_{a} Y^{a}_{ \alpha \beta} (\Psi^{\prime}_{Q})_{(i_{6}) \beta
L}
+ i  g_{sp} A^{\mu}_{j} T^{(j)}_{i_{6} j_{6}} (\Psi^{\prime}_{Q})_{(j_{6})
\alpha L}\\
             &   & +\, i \frac{g_{1}}{2} B^{\mu} y (\Psi^{\prime}_{Q})_{(i_{6})
\alpha L}
\end{array}
\label{eq12*p*ApA-1}
\end{equation}

\begin{equation}
 \cal D \mit ^{(2) \mu} (\Psi_{u})^{I}_{\alpha rt} \equiv \partial^{\mu}
(\Psi_{u})^{I}_{\alpha rt}
+ i g_{3} G^{\mu}_{a} \rm \bar{\mit Y}^{\mit a}_{\alpha \beta} \mit
(\Psi_{u})^{I}_{\beta rt}
                                       + i \frac{g_{1}}{2} B^{\mu} y
\delta_{\alpha \beta} (\Psi_{u})^{I}_{\beta rt}
\end{equation}

\begin{equation}
 \cal D \mit ^{(2) \mu} (\Psi_{d})^{I}_{\alpha rt} \equiv \partial^{\mu}
(\Psi_{d})^{I}_{\alpha rt}
+ i g_{3} G^{\mu}_{a} \rm \bar{\mit Y}^{\mit a}_{\alpha \beta} \mit
(\Psi_{d})^{I}_{\beta rt}
                                       + i \frac{g_{1}}{2} B^{\mu} y
\delta_{\alpha \beta} (\Psi_{d})^{I}_{\beta rt}
\end{equation}

\begin{equation}
(\cal D \mit _{\mu} \lambda_{G})_{a} = \partial_{\mu} \lambda_{Ga} - g_{3}
f_{abc} G_{\mu b} \lambda_{G c}
\end{equation}

\begin{equation}
W_{G \eta} = - \frac{1}{4} \bar{D} \bar{D} e^{- 2 g_{3} Y^{a} \cal G \mit _{a}}
D_{\eta} e^{2 g_{3} Y^{b} \cal G \mit _{b}},
   \;\;\;\; \rm for \;\; \mit SU_{C}(\rm 3)
\label{eq5*p*ApA-2}
\end{equation}

\begin{equation}
W_{A \eta} = - \frac{1}{4} \bar{D} \bar{D} e^{- 2 g_{sp} T^{(i)} \cal A \mit
_{i}} D_{\eta} e^{2 g_{sp} T^{(i)} \cal A \mit _{i}},
   \;\;\;\; \rm for \;\; \mit Sp_{L}(\rm 6)
\end{equation}

\begin{equation}
W_{B \eta} = - \frac{1}{4} \bar{D} \bar{D} e^{- \hat{\cal B}} \mit D_{\eta}
e^{\rm \hat{\cal B}} = \mit - \frac{1}{4} \rm \bar{\mit D} \bar{\mit D}
\mit D_{\eta} \rm \hat{\cal B},
   \;\;\;\; \rm for \;\; \mit U_{Y}(\rm 1)
\label{eq7*p*ApA-2}
\end{equation}

\end{document}